\documentclass{aa}  

\usepackage{graphicx}
\usepackage{txfonts}
\usepackage{xcolor}
\usepackage{soul}
\usepackage{float}

\def\lsim{\mathrel{\rlap{\lower3.5pt\hbox{\hskip0.5pt$\sim$}}
    \raise0.5pt\hbox{$<$}}}                
\def\gsim{~\rlap{$>$}{\lower 1.0ex\hbox{$\sim$}}}

\begin{document}

   \title{Galaxies at the edges: a complete census of MACS $J0416.1 - 2403$ cluster }


    \author{R. Ragusa
          \inst{1}\thanks{email: rossella.ragusa@inaf.it}, M. D'Addona \inst{1,2,3}, A. Mercurio \inst{1,2,4},  M. Longhetti \inst{5}, M. Girardi \inst{6,7}, M. Annunziatella \inst{8}, N. Estrada \inst{9},  C. Grillo \inst{10,11}, A. Iovino \inst{5}, G. Rodighiero \inst{12,13}, P. Rosati \inst{3,14}, B. Vulcani \inst{12}, G. Angora \inst{1}, H. B\"ohringer \inst{15,16,17}, M. Brescia \inst{18,1},   G. Caminha \inst{17,19}, G. Chon \inst{15,16}, F. Getman \inst{1}, A. Grado \inst{20,21}, M.  Gullieuszik \inst{12}, L. Limatola \inst{1}, A. Moretti \inst{5} and L. Pecoraro \inst{1,2}}
  
         \institute{\centering \textit{Affiliations can be found at the end of the document.}}    
   \date{}

 
  \abstract
   {Numerous studies have established that its surrounding environment profoundly influences the physical properties of a galaxy. While gas inflows can supply the necessary fuel for star formation, high-density and -temperature conditions can suppress star-forming activity through various quenching processes. Investigations into large-scale structures, such as filaments and overdense regions in the cluster outskirts at R$\geq$2$R_{200}$, have predominantly focused on the low-z Universe. To move to intermediate-z and explore galaxy pathways, combined with environmental effects, it is crucial to join wide-field spectroscopy and deep photometry.}
   {Our primary objective is to spectroscopically analyze the photometric overdensity structures observed by \citet{estrada2023A&A...671A.146E} in the outskirts of the massive cluster MACS J0416.1-2403 (z=0.397), interpreted as evidence of ongoing group infall into the cluster. This study aims to enhance our understanding of the evolutionary processes occurring within these substructures and their role in the pre-processing scenario. Additionally, we aim to investigate the global behavior of galaxies in the outskirts in relation to their $g-r$ color, K-band luminosity (a proxy for stellar mass) and local density, emphasizing the influence of the environment on galaxy evolution.}
   {We conducted a spectroscopic analysis extending to the outskirts up to 5.5$R_{200}$ ($\sim$10 Mpc), using the AAOmega spectrograph. The large field of view (1 deg$^2$) and depth of the observations allowed us to explore galaxies up to the cluster's periphery and across a wide stellar mass range, reaching down to the limit of dwarf galaxies. Redshifts were obtained through independent but comparable methods: Redrock, {\it EZ}, and {\it Redmost}, ensuring consistency and accuracy in our measurements.}
   {We identified 148 new spectroscopic cluster members from a sample of 1236 objects. We found that 81 out of 148 galaxies are located in filamentary and overdense regions, supporting the role of filamentary infall in the cluster mass assembly history. A spectral analysis revealed that galaxies in high-density regions are more massive, redder, and more passive, compared to galaxies in low-density regions that appear to be bluer, less massive, and more star-forming. These findings underscore the significance of environmental effects, particularly in overdense regions, and the role of pre-processing phenomena in shaping galaxy properties before cluster infall.}
   {}

   \keywords{Galaxies: evolution - Galaxies: general - Galaxies: clusters: general - Galaxies: distances and redshifts - Galaxies: clusters: intracluster medium - Galaxy: stellar content - Catalogs - Surveys - Methods: observational}
\authorrunning{R. Ragusa et al.}
\titlerunning{The galaxies at the edges: a complete census of MACS J$0416.1 -2403$ cluster}
\maketitle
%

\section{Introduction}\label{intro}

\begin{figure*}
	\includegraphics[width=18.35cm]{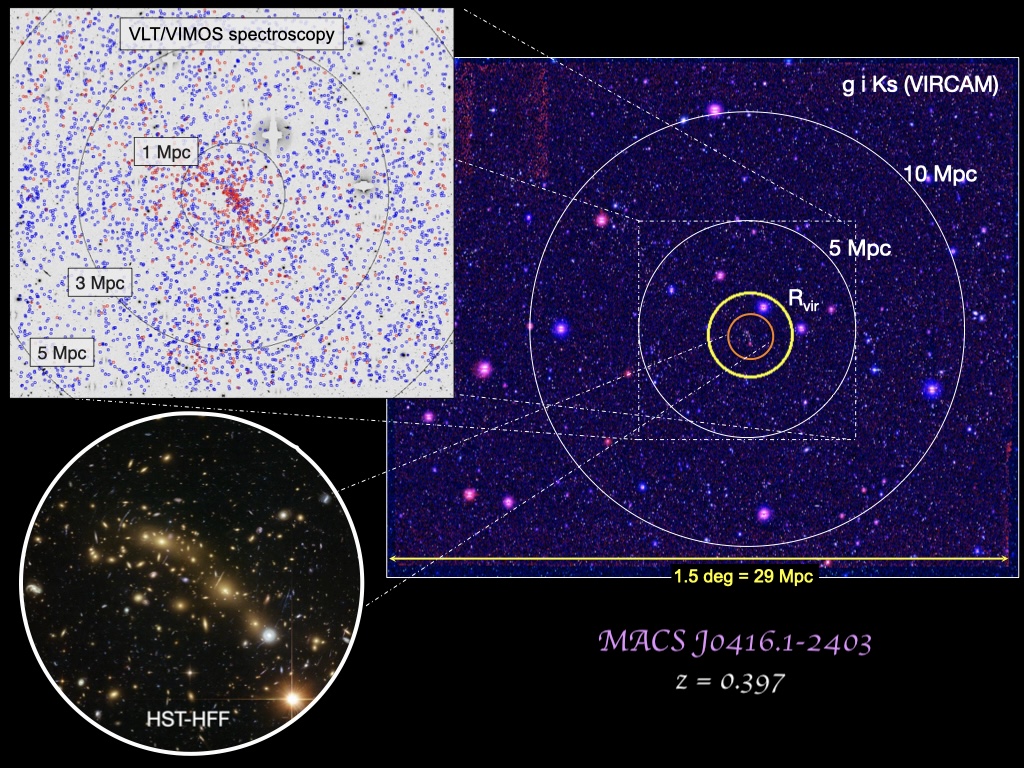}
    \caption{Imaging and redshifts spatial distribution for MACS0416. On the top left, the VST i-band image is shown, covering the 30x30 arcmin$^2$ cluster region showing MUSE (red) and VIMOS (blue) redshifts, over the range z=[0.02-6.2]. The bottom left image shows the Hubble Frontier Fields (HFF) color image (2.8' across), where MUSE integral field spectroscopy yields additional redshifts in the cluster core. The right color composite image combines VST g- and i-band plus VISTA Ks-band.}
    \label{fig:field}
\end{figure*}

In the most widely accepted cosmological model, the $\Lambda$-Cold Dark Matter scenario ($\Lambda$CDM), the formation and evolution of the structures occur hierarchically. In this framework, galaxy clusters, which are the largest gravitationally bound structures known in the Universe, are expected to grow and evolve
over time via sequential mergers and accretion of material from the surrounding structures in their outer regions \citep[e.g.,][]{Kauffmanna1999MNRAS.303..188K,Kauffmannb1999MNRAS.307..529K,DeLucia2006MNRAS.366..499D}. 
The evolution of these rich and extended structures proceeds by assembling both galaxy groups and individual galaxies. Galaxies accreted at the outskirts of clusters encounter the intracluster medium (ICM) and interact with such a dense environment. 

Thus, galaxy clusters are ideal laboratories for studying the impact of accretion through the outer regions on galaxy evolution.
One of the key unresolved questions in galaxy evolution is understanding how, when, and where galaxy properties transform.  The population of infall galaxies that interact with the hot, ionized gas of the ICM, is subjected to 
external processes, involve a large variety of mechanisms, including ram pressure stripping \citep[e.g.,][]{gunn1972ApJ...176....1G}, strangulation \citep[e.g.,][]{Larson1980ApJ...237..692L,Balogh2000ApJ...540..113B}, galaxy-galaxy interactions \citep[e.g.,][]{mihos1996ApJ...464..641M,moore1998ApJ...495..139M},  tidal stripping \citep[e.g.,][]{Zwicky1951PASP...63...61Z,Gnedin2003ApJ...582..141G,Villalobos2014MNRAS.444..313V} and thermal evaporation \citep[e.g.,][]{Cowie1977Natur.266..501C}.
Any removal or consumption of gas typically results in a reduction or inhibition of star formation, leading to galaxy quenching. Consequently, cluster galaxies exhibit different properties compared to field galaxies, although field galaxies can be also quenched \citep[e.g.,][]{Dressler1980ApJ...236..351D,Poggianti1999ApJ...518..576P}.
Since the properties of the ICM change from the cluster core to the outskirts \citep{nagai2011ApJ...731L..10N,Ichikawa2013ApJ...766...90I,lau2015ApJ...806...68L,biffi2018MNRAS.476.2689B,Mirakhor2021MNRAS.506..139M}, the efficiencies of all these processes are inhomogeneous in the intracluster space, then the galaxies in the outer regions are expected to interact differently with the ICM, showing different properties. Moreover, in the outskirts of clusters, structures resembling galaxy groups that have already transformed due to dense environments can be observed before they experience infall into the cluster as a group, a phenomenon known as pre-processing \citep[e.g.,][]{Zabludoff1998ApJ...496...39Z,Mulchaey1998ApJ...496...73M,Mihos2004cgpc.symp..277M,Fujita2004PASJ...56...29F, ragusa2021A&A...651A..39R,ragusa2022FrASS...952810R,ragusa2023A&A...670L..20R}.\\
Massive galaxy clusters, such as the target of this work, MACS $J0416.1-2403$ (hereafter MACS0416), provide an exceptional environment for studying these processes due to their rich and diverse galaxy populations since it offers a unique opportunity to study environmental effects and the collisional behavior of both luminous and dark matter during cluster formation.  
Previous studies of galaxy populations infalling into clusters have largely focused on low-redshift systems \citep{Paccagnella2016ApJ...816L..25P,Paccagnella2017ApJ...838..148P,Paccagnella2019MNRAS.482..881P,Salerno2020MNRAS.493.4950S}, and it remains crucial to extend these observations to higher redshifts. 
Despite significant advancements in our understanding, there remains ongoing debate about the exact evolutionary trajectories of galaxies and the importance of different quenching mechanisms both mass-related processes (like feedback from Active Galactic Nuclei or supernovae) and environmental factors (such as ram pressure stripping, tidal forces, harassment, collisions between galaxy groups and clusters, and gas depletion). These mechanisms likely vary in impact, depending on the redshift and environment in which they occur \citep[e.g.,][]{Annunziatella2016A&A...585A.160A}.
\begin{figure*}
 \includegraphics[width=18.7cm]{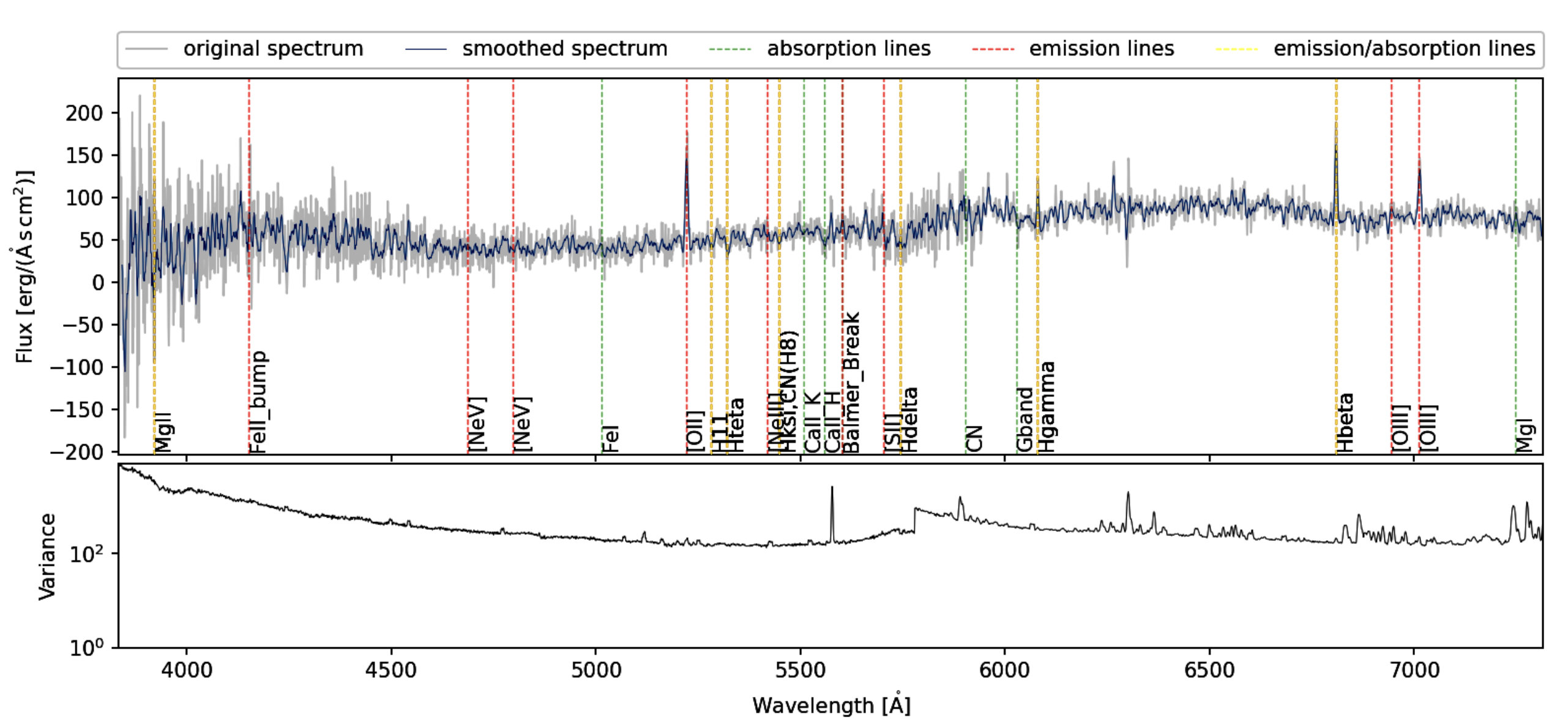}
    \caption{ {\it Top Panel}: AAOmega spectrum of a galaxy in our sample (ID: MACS$\_$64052446, z$\sim$0.40), with the z estimated by using Redrock. Key emission and absorption lines are marked with dashed lines, following the color scheme in the legend. {\it Bottom Panel}: Variance of the spectrum.}
    \label{fig:spectra}    
\end{figure*}
To fully tackle this issue, it is crucial to observe galaxy populations across a wide range of environments, from those falling into the cluster to those in more isolated regions. This includes a diversity of halo masses and capturing snapshots of galaxy evolution during periods of rapid transformation. Such a comprehensive approach will provide deeper insight into the interplay between internal and external forces driving galaxy evolution at various stages.
Despite the wealth of data on the central regions of galaxy clusters, our understanding of the physical processes taking place in the outermost regions, up to several virial radii, remains incomplete. The accretion through the outskirts is still an under-explored domain \citep{dekel2009Natur.457..451D,Danovich2012MNRAS.422.1732D,walker2019SSRv..215....7W,Welker2020MNRAS.491.2864W}. This is partly due to the observational challenges in probing the lower-density outer regions of galaxy clusters, where the ICM is less dense and thus harder to detect in X-ray and Sunyaev–Zeldovich observations. Furthermore, most studies focus on the central regions, covering only a fraction of the ICM volume. At the same time, the outskirts can reveal critical information about gas accretion, dark matter distribution, and galaxy transformation mechanisms beyond the dense cluster core.
At intermediate redshifts (z$\sim$0.4), the lack of comprehensive observational campaigns targeting the cluster outskirts beyond 2-3 virial radii leaves a significant gap in our understanding of mass assembly processes. Observing galaxies out to 5$ R_{200}$ is crucial, as this is where galaxies undergo pre-processing in group environments before falling into the cluster \citep{Fujita2004PASJ...56...29F}. The outer regions also offer key insights into dark matter dynamics, including the role of caustics in the dark matter density profile \citep{Mansfield2017ApJ...841...34M,diemer2017ApJ...843..140D} and the contribution of turbulent gas motions \citep{lau2009ApJ...705.1129L,vazza2009A&A...504...33V,Battaglia2012ApJ...758...74B}. Furthermore, characterizing the gas dynamics in these regions can shed light on non-equilibrium processes affecting the ICM, such as bulk and turbulent gas flows \citep{rudd2009ApJ...701L..16R} and an inhomogeneous gas density distribution \citep{nagai2011ApJ...731L..10N,Roncarelli2013MNRAS.432.3030R}.
This work aims to expand the existing Multi Unit Spectroscopic Explorer (MUSE) and VIsible Multi-Object Spectrograph (VIMOS) data \citep[CLASH-VLT;][]{rosati2014Msngr.158...48R} by targeting, with the 2dF+AAOmega instrument, a larger region of MACSJ0416 up to 5.5$R_{200}$ ($R_{200}$=1.82 Mpc), building on the low- and medium-resolution data previously collected for $\sim$4700 galaxies, of which $\sim$ 950 are confirmed cluster members. Complementary to these data, further observations from the cluster's core, such as the Grism Lens-Amplified Survey from Space \citep[GLASS;][]{treu2015ApJ...812..114T, vulcani2016ApJ...833..178V,Vulcani2017ApJ...837..126V} spectroscopic data and archival MUSE integral-field spectrometry \citep{Caminha2017A&A...600A..90C}, will enhance the study's depth. This campaign will integrate multiwavelength observations, including optical data from VLT Survey Telescope (VST) \citep{estrada2023A&A...671A.146E}, NIR data from Visible and Infrared Survey Telescope for Astronomy (VISTA), X-ray, and radio datasets, to provide a well-rounded view of galaxy populations within the cluster up to the periphery. The comprehensive dataset will include galaxies down to the dwarf regime (stellar mass $ \sim 10^9 M_{\odot}$), allowing for a complete exploration of galaxy evolution from the cluster core to well beyond the virial radius.
The paper is organized as follows.
In Sec. \ref{target}, we describe the state-of-the-art on MACS0416 at the time of this work. We illustrate the observations, the data reduction, and the data analysis in terms of redshift
measurements, in Sec. \ref{obs}.
In Sec. \ref{catalogs}, we report the AAOmega spectroscopic catalog and the criteria for the member selection. In Sec. \ref{results}, we present the results: in Sec. \ref{redvsblu}, Sec. \ref{kmag} and Sec. \ref{Overdens} we qualitatively analyze the spectral properties of the galaxies in our sample of members respect with the $g-r$ color, the magnitude in the K-band as proxy of the stellar mass and properties of the galaxies within the different density range in the field of view (FoV), respectively.
Finally, in Sec. \ref{conclusions} we
summarize our results and conclusions.\\
Throughout the paper, we adopt the ($\Lambda$CDM)  cosmology with $\Omega_m$=0.3,
$\Omega_\Lambda$=0.7, and $H_0$=70 km s$^{-1}$ Mpc$^{-1}$. Therefore, 1 arcmin corresponds to $\sim$0.321 Mpc at z=0.397.
All magnitudes are given in the AB system.

\section{MACS J0416.1-2403}\label{target}

MACS0416 is a massive galaxy cluster located at redshift z=0.397$\pm$0.001 ( $\sim$4.26 Gyr, i.e., 30$\%$ of the lookback time to the Big Bang)\citep{Balestra2016ApJS..224...33B}, particularly known for its strong lensing effects and its complex structure. Fig. \ref{fig:field} shows images of MACS0416 obtained using all the photometric data available at the time this work was written. Galaxies for which spectroscopic redshifts were obtained with MUSE (in red) or VIMOS (in blue) are marked in the top left panel. The color composite image of HST FoV is illustrated in the bottom left panel, while the one from VST g- and i-bands plus VISTA Ks-band is shown in the right panel.  The cluster, originally identified as part of the MACS survey \citep{Ebeling2001ApJ...553..668E}, is characterized by high X-ray luminosity L$_X$ (L$_{500}$)=(1.62 $\pm$ 0.02)$\times10^{45}$ L$_{\odot}$ \citep{Ogrean2016ApJ...819..113O} and by a very high total mass of approximately 
M$_{200}$=1.04$\times$10$^{15}$ M$_{\odot}$ \citep{UMETSU2014ApJ...795..163U}. \citet{Mann&Ebeling2012MNRAS.420.2120M} classified MACS0416 as a merger system due to its irregular X-ray morphology and the projected separation of roughly 
200 kpc between the two brightest cluster galaxies (BCGs).  In support of this, the weak and strong lensing analyses \citep{zitrin2013ApJ...762L..30Z,Jauzac2014MNRAS.443.1549J,grillo2015ApJ...800...38G,Jauzac2015MNRAS.446.4132J,Richard2014MNRAS.444..268R,hoag2016ApJ...831..182H,Caminha2017A&A...600A..90C,Bonamigo2017ApJ...842..132B,Bonamigo2018ApJ...864...98B,Chirivi2018A&A...614A...8C,Bergamini2023A&A...674A..79B} revealed an elongated projected total mass distribution in the cluster core, with two central mass concentrations (a characteristic typical of merging clusters), along with potential secondary structures located to the southwest and northeast, both approximately 2 arcmin from the cluster's center.  
Recognized as one of the most potent gravitational lenses in the universe \citep{vanzella2021A&A...646A..57V,mestric2022MNRAS.516.3532M,Bergamini2023A&A...674A..79B}, MACS0416 was initially imaged by the Hubble Space Telescope (HST) as part of the Cluster Lensing And Supernova survey with Hubble \citep[CLASH;][]{postman2012ApJS..199...25P}. Subsequent observations under the HFF initiative \citep{lotz2017ApJ...837...97L} produced deep imaging data, achieving a point-source detection limit of approximately 29 AB-mag.

\begin{table}
\setlength{\tabcolsep}{4pt}
\centering
\caption{New catalog of spectroscopic redshift of MACS0416 field obtained with AAOmega observations.} 
\begin{tabular}{lccccc}
\hline\hline
Target ID & RA [deg]& DEC [deg]& z &  QF \\
\hline
\\ 
MACS0416$\_$63922461	&63.925583	&$-24.612942$	&2.760	&3\\
MACS0416$\_$64012446	&64.014958	&$-24.465181$	&0.399&	3\\
MACS0416$\_$64522446	&64.526917	&$-24.462817$	&0.297&	3\\

.......& 	& &	 	& \\
\hline\hline
\label{TAB: CATALOG}

\end{tabular}
\tablefoot{Col. 1 contains the object ID; In Col. 2 and 3 are listed the RA and DEC; Col. 4 includes the redshift estimated (z) and Col. 5 their QF (see Sec. \ref{catalogs}).}
\end{table}

Within HFF investigation, with a high-resolution dissection of the two-dimensional total mass distribution in the core of MACS0416, \citet{Annunziatella2017ApJ...851...81A} found no significant offset between the stellar and dark matter components within the core of the cluster.
Moreover, \citet{Bonamigo2017ApJ...842..132B} quantified the gas-to-total mass fraction at approximately 10$\%$ at a distance of 350 kpc from the cluster center, demonstrating that the dark matter to total mass fraction remains relatively stable up to that distance. 
Recently, MACS0416 has been observed by the James Webb Space Telescope (JWST) as part of the Prime Extragalactic Areas for Reionization and Lensing Science \citep[PEARLS;][]{Windhorst2023AJ....165...13W}. These observations extended the wavelength range to approximately 5$\mu$m, offering enhanced resolution and sensitivity in the infrared compared to previous HST data. 
MACS0416 is also a focal point of the European Southern Observatory's (ESO) Large Programme on "Dark Matter Mass Distributions of Hubble Treasury Clusters and the Foundations of $\Lambda$CDM Structure Formation Models" \citep[CLASH-VLT;][]{rosati2014Msngr.158...48R}, utilizing the VIMOS data at the Very Large Telescope (VLT). In their study, \citet{Balestra2016ApJS..224...33B}, presenting the large spectroscopic campaign that provided $\geq$ 4000 reliable redshifts over $\sim$600 arcmin$^2$ including $\sim$800 cluster member galaxies, confirmed the complex dynamical structure of the cluster, characterized by a double-peaked profile suggestive of a merging cluster, supporting the notion that the system is observed in a pre-collisional state, consistent with findings from radio and deep X-ray analyses by \citet{Ogrean2015ApJ...812..153O}. The velocity dispersion for each of these components was found to be $\sigma_1\sim 799_{-20}^{+22}$ and $\sigma_2\sim 955_{-22}^{+17}$ km s$^{-1}$ \citep{Ebeling2014ApJS..211...21E,Jauzac2014MNRAS.443.1549J}. 
In light of the above, the combination of photometric and spectroscopic data now available for MACS0416, derived from extensive HST, VLT, Chandra and JWST observations, places it as one of the best datasets for investigating the dark matter distribution in a massive merging cluster through strong lensing techniques \citep{diego2024A&A...690A.114D}. 
Furthermore, MACS0416 also serves as an ideal laboratory for studying stellar populations and the effects of the environment on galaxy populations within massive clusters.
Within the CLASH-VLT survey, \citet{Olave-Rojas2018MNRAS.479.2328O}, based on photometric and spectroscopic analysis, dynamically identified the two aforementioned substructures within a radius less than 2$R_{200}$ surrounding MACS0416. The authors also analyze galaxy colors to explore how pre-processing impacts the suppression (quenching) of star formation in these galaxies. By comparing the fractions of red (passive) and blue (star-forming) galaxies in the main cluster and its substructures, the authors observe consistent spatial trends, since in both the main cluster and substructures, galaxies further from the cluster center ($r \geq R_{200}$) show a lower fraction of blue galaxies compared to field galaxies. Additionally, the study finds that at large distances, the quenching efficiency of substructures is similar to that observed in the main cluster, suggesting that pre-processing significantly influences galaxy evolution and quenching in these environments.
The analysis highlights that pre-processing, occurring in substructures, is an important factor in the transition of galaxies into passive states at low redshifts.
Finally in a recent work by \citet{estrada2023A&A...671A.146E}, the authors provide an in-depth analysis of galaxy assembly and the influence of the environment on the galaxy population within the massive cluster MACS0416. Using data obtained from the Galaxy Assembly as a Function of Mass and Environment
with VST survey (VST-GAME survey, P.I. A. Mercurio), combined with VISTA data (G-CAV survey, P.I. M. Nonino) the study focuses on photometric analysis to explore the density field and structure of MACS0416, with particular attention to the outskirts of the cluster where signs of galaxies or galaxy groups infall are detected.
They apply photometric redshift techniques and galaxy classification methods not only to generate a robust catalog of candidate cluster members, based on their photometric properties, but also to identify galaxy populations across different environments, including the dense core and the more diffuse outer regions of the cluster. The analysis reveals that the $g-r$ color distribution exhibits a bimodal pattern across all environments,
 with red galaxies displaying a shift toward redder colors as the environmental density increases. Conversely, the fraction of blue cloud galaxies increases as the environmental density decreases. 
The key findings of the study highlight the existence of three distinct overdensity structures in the outskirts of MACS0416 at approximately $5.5 R_{200}$, suggesting ongoing accretion of galaxy groups into the cluster, a clear indication of the pre-processing scenario. The authors conclude that these structures represent infalling groups, which may significantly contribute to the galaxy population in the cluster, affecting their subsequent evolution through environmental interactions. 
In terms of galaxy properties, the study finds that the galaxies within these overdensity regions exhibit mean densities and luminosities similar to those of the cluster core and different characteristics compared to those in the field.
Their colors suggest that these galaxies may be part of evolved populations, providing further evidence for pre-processing effects acting on galaxies in these substructures.
The analysis presented in this work constitutes a spectroscopic follow-up of the photometric sample by \citet{estrada2023A&A...671A.146E}. Our main goal is to spectroscopically analyze the overdensity structures they identified in the outskirts of MACS0416, providing evidence of ongoing group infall into the cluster. This spectroscopic analysis will provide a more detailed understanding of the evolution of these substructures and their contribution to the pre-processing scenario. Additionally, we aim to qualitatively study the stellar populations as a function of $g-r$ color, K-band luminosity (as a proxy for the stellar mass), and within different environments of the cluster in terms of local density, shedding light on the influence of the cluster environment on galaxy evolution.
\section{Observations, data reduction and redshift measurement}\label{obs}

Observations were conducted over four nights, from November 2 to November 5 2021 (program ID: O/2021B/008, P.I. A. Mercurio), using the {\it Two Degree Field (2dF)} Multi-Object System (MOS) combined with the {\it AAOmega Spectrograph} mounted on the Anglo-Australian Telescope {\it (2dF+AAOmega)}.
During dark time and under clear weather conditions (seeing$_{max}$ = 1.3"), six AAOmega MOS runs, each with an exposure time of 3 hr (i.e. total exposure time $\sim$18 hr), were planned to be carried out using different fiber configurations but sharing the same center. The goal was to achieve a signal-to-noise ratio (S/N)$\geq$5 for objects with a V-band magnitude limit of V$_{lim}$<21.5 mag.
Unfortunately, out of the $\sim$2000 galaxies initially expected, only $\sim$1290 targets were observed, that is $\sim 30\%$ of the total number of objects in the photometric catalog with V$_{lim}$<21.5 mag. This shortfall was due to time lost because of bad weather, which affected two nights (i.e., 50$\%$ of the total allocated time). This incompleteness prevented us from fully spectroscopically covering all the photometric overdensities detected by \citet{estrada2023A&A...671A.146E} and partially limited the statistical significance of the sample. However, despite this limitation, our analysis remains robust and significant within the observed sample, supporting the scientific validity and relevance of our results (see Sec. \ref{results})
The {\it 2dF+AAOmega Spectrograph} data were acquired with the grating sets 385R and 580V, which cover the wavelength ranges from
370nm to 580nm ("blue" arm) and from 560nm to 950nm ("red" arm), respectively, with a resolution R=1300 \citep{Sharp2006SPIE.6269E..0GS}. 

\begin{figure}
	\includegraphics[width=9 cm]{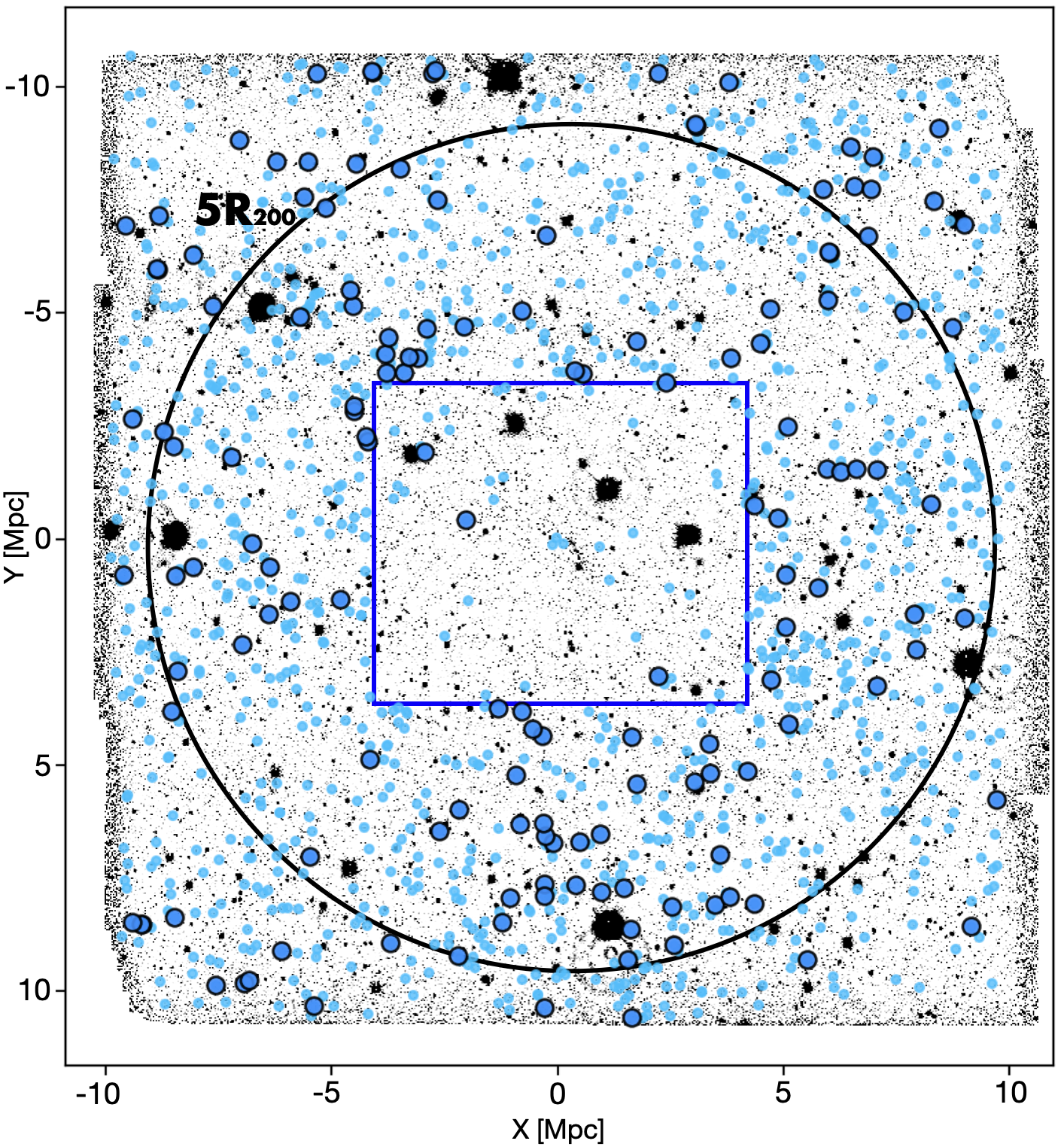}
    \caption{
VST i-band image covering 1 deg$^2$
($\sim$ 20$\times$20 Mpc$^2$), overlapped with the 1236
 AAOmega measured redshifts (turquoise points).  The light blue circles with black edges represent the new 148 members of MACS0416 from AAOmega catalog, in the redshift range z=[$0.382-0.412$]. The region where VIMOS or MUSE redshifts are available over
 z=[$0.02-6.2$] range is represented as a blue rectangle. The black circle corresponds to the radial distance from the cluster center of 5R$_{200}\sim$9.1 Mpc}
    \label{fig:iband}
\end{figure}

\begin{figure*}
	\includegraphics[width=18.5cm]{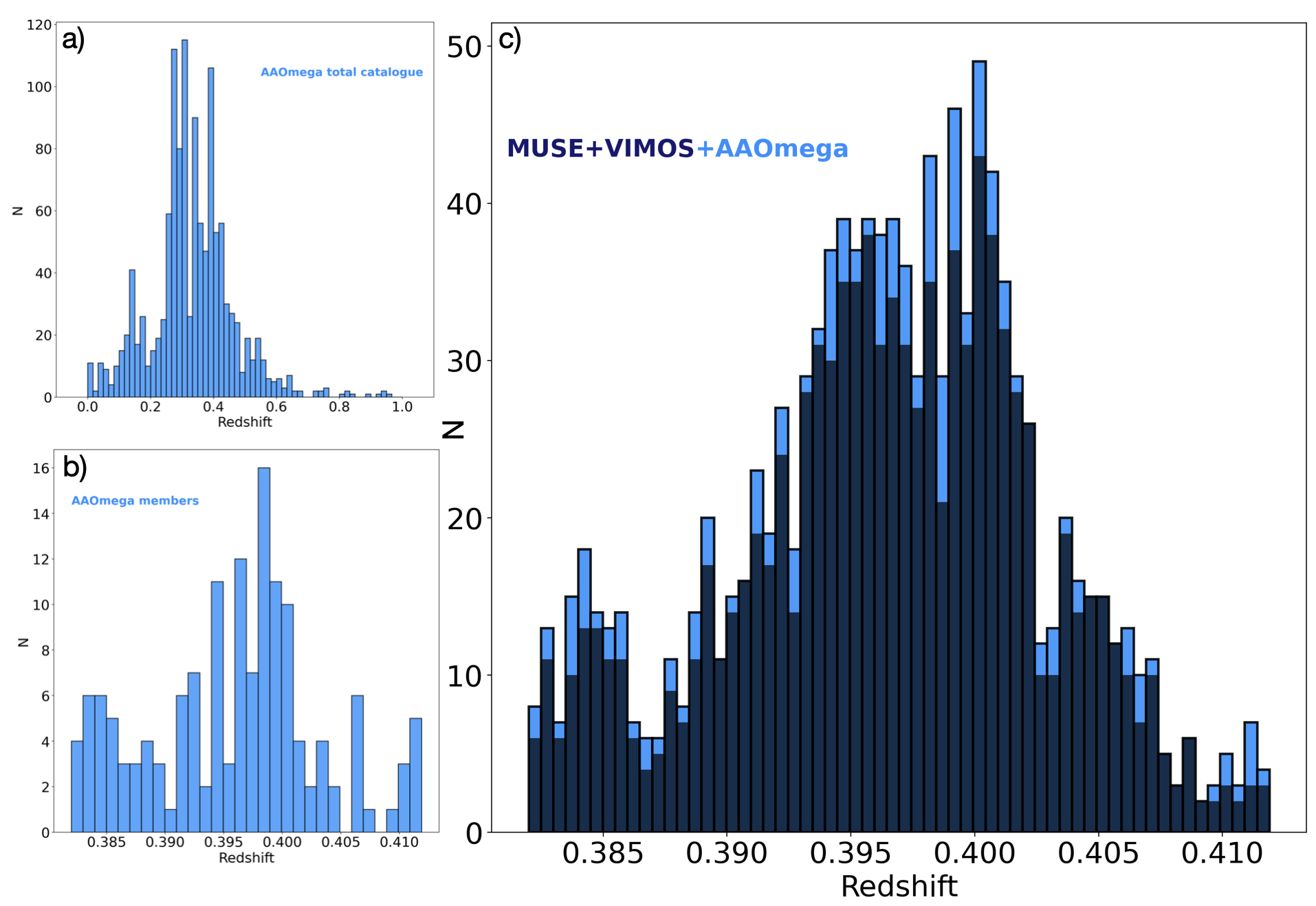}
    \caption{ {\it Top left panel}: redshift distribution of the 1233 sources extracted from the AAOmega spectra. Three objects with redshift $\geq$1 were excluded from the plot for a better visualization of the distribution; {\it Bottom left panel}: Zoom into the redshift range of the members selected with  AAOmega observations;
{\it Right panel}: Redshift distribution of the 1017 from MUSE or VIMOS (blue) plus 148  
 from AAOmega (turquoise) spectroscopic cluster members selected in the redshift range z=[0.382-0.412].}
    \label{fig:histogram}
\end{figure*}
\begin{figure*}
   
    \includegraphics[width=9.15cm]{figures/mean_spectrum_col_g_r_kron_blu_vs_red_1.15.jpg}
    \includegraphics[width=9.15cm]{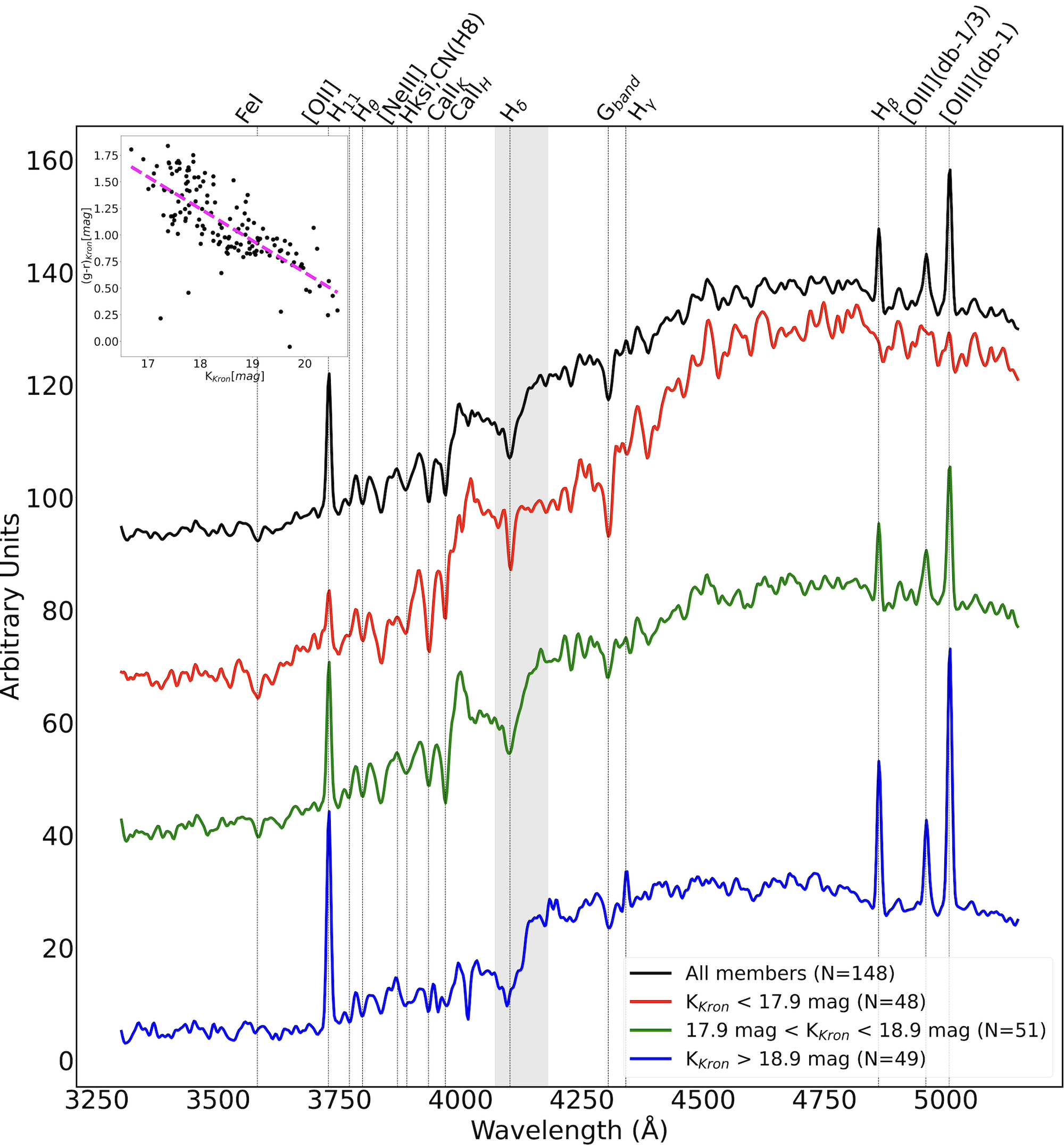}
    \caption{{\it Average spectral properties of our sample as a function of color and K$_{\mathrm{Kron}}$ as a proxy of the stellar mass.} Left panel: Using a $(g-r)_{\mathrm{Kron}}$ color cut, the sample was divided into red (72) and blue (76) galaxies, with their stacked spectra displayed in red and blue, respectively. The inset in the top left corner shows a color distribution of our sample, where the x-axis represents the observed $(g-r)_{\mathrm{Kron}}$ color, corrected for extinction. Right panel: The whole sample was divided into three subclasses based on the three bins of K$_{\mathrm{Kron}}$. In red is the stacked spectrum of the brightest galaxies with K$_{\mathrm{Kron}}$<17.9 mag. In green that of the galaxies with intermediate luminosity 17.9 mag$\leq$K$_{\mathrm{Kron}}$< 18.9 mag and in blue are the stacked spectrum of the faint galaxies with K$_{\mathrm{Kron}}$$\geq$18.9 mag. In the top left corner, an inset showing the trend between $(g-r)_{\mathrm{Kron}}$ and K$_{\mathrm{Kron}}$ magnitude is reported. The dashed magenta line corresponds to the best fit of the linear correlation. In both panels, the stacked spectrum of the entire population of members (148) of MACS0416, obtained with AAOmega, is shown in black. All spectra are rest-frame and have been smoothed using a Gaussian kernel with a sigma of 2 pixels. The vertical dashed gray area corresponds to the spectral region of the join between the two arms.}
    \label{fig:col_vs_bluredandmag}
\end{figure*}

We reduced the raw data and merged the spectra from the two arms
using the standard 2dfdr pipeline\footnote{2dfdr is an automatic data reduction
pipeline dedicated to reducing multi-fibre spectroscopy data from the AAT facility. For more
information, see https://aat.anu.edu.au/science/software/2dfdr} v5.33 with default configuration and with the Principal
Component Analysis (PCA) that enabled the improvement of sky background subtraction. It is worth noting that the sky contribution was subtracted on the basis of only eight observed sky spectra, an insufficient number to accurately sample the sky across the entire FoV and this affects mainly the redder part of the spectra. For this reason, we applied cuts to the blue and red ends of each spectrum, excluding the low S/N range in the blue and the residuals of telluric absorptions in the red range. 
After these cuts, our 1290 spectra cover a wavelength range
[3800-7300] $\AA$. We refer the reader to App. \ref{appA} for more details about the data reduction.
Among these, 24 spectra correspond to 6 galaxies with previously known redshifts, estimated using MUSE data \citep{Caminha2017A&A...600A..90C}. These galaxies were repeatedly observed in each MOS configuration for calibration purposes. This leaves us with a final sample of 1272 unique spectra, of which 1266 correspond to new, previously unobserved objects (see Sec. \ref{catalogs} for details).
Spectroscopic redshifts in this work were determined using two complementary methods: 1) an automatic redshift estimation tool, namely Redrock \citep{guy2023AJ....165..144G}; 2) visual inspection of all spectra to validate Redrock's results and manually estimate the redshift when the automatic output was deemed unsatisfactory. In the latter case, the estimation was performed using the {\it Pandora ez} software \citep{Garilli2010PASP..122..827G}. 
In App. \ref{appB}, both the independent methods are described to provide a comprehensive overview of the procedures for obtaining spectroscopic redshifts.
We also developed and implemented a new complementary software tool, Redmost \footnote{\url{https://github.com/mauritiusdadd/redmost}}, designed for redshift determination. This tool allowed us to use {\it Redrock } for automatic redshift determination while offering a graphical interface for real-time verification of the quality of the results produced by {\it Redrock }. In cases where the reliability of the automatic determination was low, and the estimation had not yet been refined with Pandora {\it EZ }, we were able to manually intervene and refine the redshift estimate similarly to {\it EZ }, as described in detail in App. \ref{appB}.
In Fig. \ref{fig:spectra}, we show the spectrum of one of the galaxies in our sample (at the average z of the cluster), where the redshift estimation by Redrock was done.
The spectroscopic catalog thus obtained is described in detail in Sec. \ref{catalogs}. 
It is worth noting that, at this stage of the analysis, no absolute flux calibration has been applied to the spectra, as this lies beyond the scope of the present work. However, for the purpose of an internal comparison between spectra aimed at showcasing the potential of our dataset, the lack of absolute flux calibration does not compromise the scientific value or potential of the sample.

\section{The outskirts of MACS0416: AAOmega spectroscopic catalog}\label{catalogs}

In this section, we present the spectroscopic catalog of the redshifts for the newly acquired objects.
These additional sources in the field further enrich the already existing spectroscopic catalog for this cluster \citep{Balestra2016ApJS..224...33B}. 
After estimating the redshifts from the extracted spectra, we proceeded with a visual inspection to assign a Quality Flag (QF) to each redshift estimate, following the same criteria as \citet{Balestra2016ApJS..224...33B} and \citet{Mercurio2021A&A...656A.147M}:
\begin{itemize}
    \item {\it QF = 1}: Uncertain redshift estimate; Low signal-to-noise ratio, which hinders the clear identification of spectral features, resulting in a reliability of $\sim$20-40$\%$;
    \item {\it QF = 2}: Redshifts are likely well estimated; Detection of at least two emission or absorption features, yielding an 80$\%$ reliability;
    \item {\it QF = 3}: A reliable redshift estimate; Multiple emission lines and/or absorption features are identified, ensuring a 100$\%$ confidence level in the redshift measurement;
    \item {\it QF = 9}: Emission line-based redshift estimation; One or (few) more emission lines are detected, leading to a reliability of $\geq$ than 90$\%$.
\end{itemize}

For 17 sources among the previously unobserved 1266 galaxies, no redshift estimate could be obtained. We successfully estimated redshifts for 1249 objects with a QF $\geq$1, resulting in a success rate of $\simeq$98.7$\%$. Within this subset, 13 objects were assigned a redshift quality flag QF=1 ($\sim1\%$ ), 165 QF=2 ($\sim13\%$ ), 1065 QF=3 ($\sim84\%$ ), and 6 galaxies were flagged with QF=9 ($\sim1\%$ ). Therefore, if we conservatively consider only those objects in our catalog with a QF$\geq$2, which totals 1236 out of the original dataset of 1266 sources, the success rate corresponds to $\simeq$97.6$\%$.
The resulting spectroscopic catalog, which is publicly available at the CLASH-VLT website \footnote{\url{https://sites.google.com/site/vltclashpublic/data-release}}, includes the information, as listed in Tab. \ref{TAB: CATALOG}, for the galaxies for which the redshift estimation is characterized by a QF$\geq$2.
%
%
The spatial distribution of all objects included in the complete AAOmega catalog, is shown in Fig. \ref{fig:iband} overlaid on the wide-field i-band image from the VST telescope. Specifically, all these 1236 objects with redshifts measured using AAOmega are represented in turquoise. These objects complement an existing catalog of 4711 sources in the MACS0416 field, with redshifts previously estimated using VIMOS or MUSE instruments, limited to within $\sim$ 2$R_{200}$. This VIMOS/MUSE region is shown in the figure with a blue rectangle. The AAOmega members (see Sec. \ref{spat_distrib}) are shown as light blue circles with black edges. As shown in Fig. \ref{fig:iband}, it is worth highlighting that the AAOmega dataset allows us to probe cluster regions extending well beyond 5$R_{200}$, corresponding to distances greater than 10 Mpc from the cluster center. This extensive coverage provides the necessary means to achieve our primary goal: analyzing the global properties of galaxies in the outskirts of MACS0416. 
The spectroscopic redshift distribution of the MACS0416 cluster field
from the AAOmega catalog is shown in the (a) panel of Fig. \ref{fig:histogram}. 
Only objects with redshift z $\leq$ 1 (1233 out of 1236) are displayed to enhance the visualization of the statistical distribution of sources around the redshift (z=0.397) of the cluster. Three galaxies 
have z $\geq$ 1, with values of z= 1.38, 2.76, 4.32.
In the next section, we provide the selection of the galaxies in our catalog that belong to the cluster. 
A statistically significant peak of objects around the cluster redshift (z=0.397) is evident in the distribution, indicating the detection of a significant number of objects belonging to the cluster between $\sim$2$R_{200}$ and $\sim$5.5$R_{200}$ in our AAOmega spectroscopic catalog. These objects populate the cluster outskirts. Another significant peak is detected at z $\sim$ 0.30, likely representing a foreground structure that serendipitously falls within our FoV. 

\subsection{Selection of cluster members and their redshift distibution}\label{spat_distrib}

In this Section, we perform the selection of the spectroscopic cluster members.
The dynamical analysis of the cluster members is the subject of a forthcoming paper (Ragusa et al., in prep.), and it is beyond the scope of this study. Here, we focus on analyzing the global properties of galaxies selected as spectroscopic  fiducial members of the cluster adopting the redshift range z = [0.382–0.412], corresponding to $\Delta$z = 0.015, derived from the fiducial velocity range $\pm$3000 km/s (i.e. $\Delta$z = 0.014, used in \citet{Balestra2016ApJS..224...33B, rosati2014Msngr.158...48R} and accounting for the uncertainty in the cluster's redshift center ($\Delta$z = 0.001). Between z = 0.382 and z = 0.412, we identified a sample of 148 new members that populate the outskirts of MACS0416. Hereafter, the term {\it cluster members} refers generically to galaxies within the redshift range z = [0.382–0.412], including different populations such as potential infalling or backsplash galaxies, since a detailed dynamical analysis will be presented in a forthcoming work. The redshift distribution of these new objects, shown in the (b) panel of Fig. \ref{fig:histogram}.  We recover the same statistically significant redshift peaks previously identified in the VIMOS sample (Balestra et al. 2016), with the AAOmega data independently reinforcing the robustness and reliability of these structures. The highest peak is around the cluster's mean redshift (z = 0.397). Additionally, smaller peaks, such as the one around z = 0.384, are also observed. These may correspond to the overdense regions (i.e., groups undergoing infall, namely pre-processing) that we aim to identify (see Sec. \ref{Overdens}).
In the (c) panel of Fig. \ref{fig:histogram}, the statistical distribution of the entire redshift sample available in the MACS0416 field (a total of 1165 objects) within the redshift range z = [0.382–0.412] is shown. This distribution was obtained by expanding the previously observed sample (in blue) with the addition of 148 new reliable AAOmega member galaxies (in turquoise).
Moreover, the secondary peaks around z$\sim$0.384 and z$\sim$0.405, already partially visible in the redshift distribution of the AAOmega sample, appear even more pronounced in the total catalog MUSE+VIMOS+AAOmega.
In the following section, we analyze the member galaxies' global features from the AAOmega spectroscopic dataset to characterize their average physical properties.

\begin{figure*}
	\includegraphics[width=9.5cm]{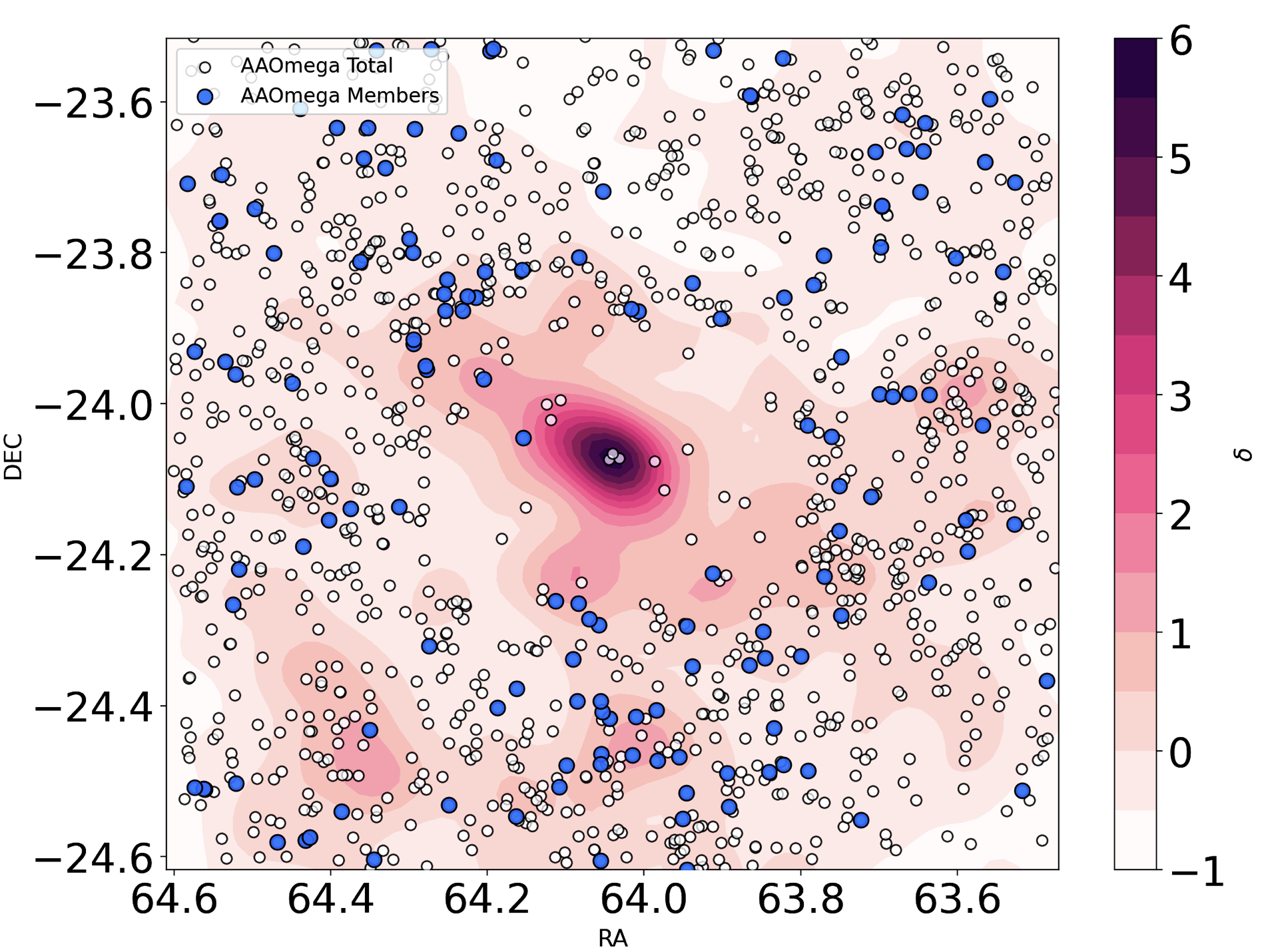}
    \includegraphics[width=9.45cm]{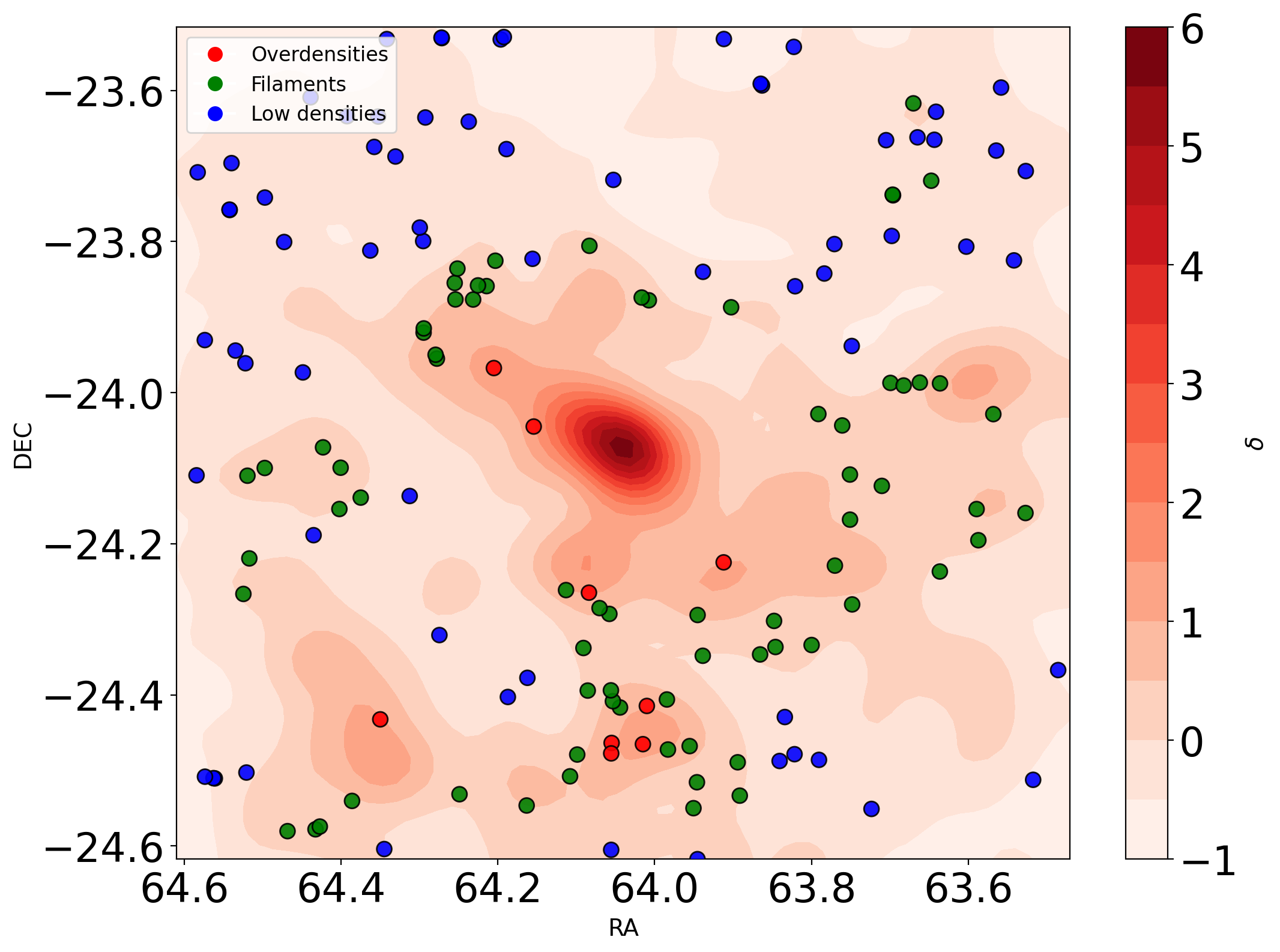}
    \caption{Left Panel: Photometric density map of MACS0416 from \citet{estrada2023A&A...671A.146E}, with all the new redshifts of objects with QF$\geq$2 (1236) obtained with AAOmega overlaid (white circles), highlighting the new spectroscopic cluster members (148) (light blue circles).  Right Panel: Same as left panel, with all the new spectroscopic cluster members (148)  obtained with AAOmega overlaid, color coded as a function of $\delta$: red and green points are the members in the overdensities plus filamentary structures, while blue points are the ones in the underdense regions.}
    \label{fig:overdensities}
\end{figure*}
\section{A complete census of MACS0416}\label{results}

In this work, we present the 2dF+AAOmega spectroscopic dataset of the galaxy cluster MACS0416 at z=0.397, focusing on its peripheral regions. 
This enabled the study of spectral properties' relationships with the cluster environment, including the detection of overdensities and filamentary regions,  photometrically detected by \citet{estrada2023A&A...671A.146E}.
The available optical and NIR catalogs enable a comprehensive characterization of galaxies down to the dwarf scale (M$_\star \sim 10^9 M_\odot$).
These combined data-sets will serve as an invaluable resource for studying galaxy evolution across a broad range of environments, from the core of the cluster to its outskirts at $\sim$5.5$R_{200}$ and offer an opportunity to explore galaxy properties and mass assembly at intermediate redshift (z$\sim$0.4), providing insights that cannot be matched by studies of other clusters at similar redshifts. 

\subsection{The tale of red and blue galaxies: color dependence of galaxy properties}\label{redvsblu}

Building upon the work in Sec. \ref{spat_distrib}, we match our sample to the \citet{estrada2023A&A...671A.146E} photometric catalog and use the galaxy $(g-r)_{\mathrm{Kron}}$ color (i.e. the $g-r$ color inside an aperture of one Kron radius) to separate blue and red galaxies. We adopt the color cut founded by \citet{estrada2023A&A...671A.146E}, i.e. $(g-r)_{\mathrm{Kron}}$$\sim$1.20 mag. We also correct the magnitudes for the Galactic extinction using the extinction coefficients
provided by \citet{Schlafly2011ApJ...737..103S}. In this way, we obtain 72 red and 76 blue galaxies.
At this early stage, the classification relies solely on the observed $(g-r)_{\mathrm{Kron}}$ color, without yet incorporating any spectral property.
As a first qualitative result of this analysis, and as evident from this simple color cut,  we observe that, contrary to expectations \citep[the so-called morphological segregation or morphology-density relation; e.g.,][]{Oemler1974ApJ...194....1O,Dressler1980ApJ...236..351D,vulcani2023ApJ...949...73V}, the number of blue galaxies is not significantly higher than that of red galaxies. Instead, both populations appear to be comparable in number within our sample. This seems to suggest that we are likely observing galaxies that reside in more evolved structures in the outskirts, tracing regions of higher environmental density in the periphery of MACS0416. 
In the left panel of Fig. \ref{fig:col_vs_bluredandmag} we show the average spectral properties of our sample as a function of color: in black is the average spectrum of all 148 AAOmega members of MACS0416, in red the one of all the 72 red galaxies and in blue the average spectrum of all the 76 blue galaxies, following the color cut described above.
All the spectra have been shifted to the rest frame and smoothed with a Gaussian kernel with a sigma of 2 pixels. Moreover, the color distribution of our sample is illustrated in the inset in the top left corner of left panel of Fig. \ref{fig:col_vs_bluredandmag}, where the x-axis represents the observed $(g-r)_{\mathrm{Kron}}$ color corrected for extinction.  We have color-coded the galaxies in the distribution according to the color cut defined above, specifically highlighting the distribution of blue galaxies (in blue) and red galaxies (in red).
As expected, red and blue galaxies show significant spectral differences due to their varying star formation (SF) activities. Red galaxies are typically old, passive systems with little to no ongoing star formation (the so-called Early Type Galaxies, ETGs). These galaxies tend to exhibit prominent absorption features in their spectra from heavy elements, like calcium (Ca, particularly prominent is the calcium $\rm{CaII}_{\rm{H}}$ and $\rm{CaII}_{\rm{K}}$ absorption lines doublet), iron (Fe), magnesium (Mg), etc., reflecting the contribution of old star populations. Mostly prominent in passive galaxies is then the Balmer break, a discontinuity in the spectral continuum, seen around $4000~\AA$ in the rest-frame, that can be used as a key indicator of the age of the stellar population in a galaxy  \citep[][]{Bruzual2003MNRAS.344.1000B, Kauffmann2003MNRAS.341...33K,Gallazzi2005MNRAS.362...41G,Franzetti2007A&A...465..711F}.
In contrast, blue and star-forming galaxies, also known as Late-Type Galaxies (LTGs), are actively forming stars, and are characterized by prominent emission lines in the spectrum, such as the [OII]$\lambda$3727, [OIII]$\lambda$5007-4959 doublet, $\rm{H}_{\beta}$, $\rm{H}_{\alpha}$. 
These differences in the galaxy populations are crucial for understanding galaxy evolution, as they reflect different stellar ages, different star formation histories, and different gas content in galaxies, and they serve as important indicators for classifying galaxies and understanding their role within the cosmic web of large-scale structures.
In our sample, the stacked spectrum of the entire AAOmega sample (in black, left panel of Fig. \ref{fig:col_vs_bluredandmag}) reveals distinct features indicative of a young stellar population, reflecting mostly the features of the blue galaxies (in blue, left panel of Fig. \ref{fig:col_vs_bluredandmag}) such as the strong emission lines of [OII], [OIII] and H$_{\beta}$, which are consistent with galaxies typically found in the outer parts of clusters or field. Such galaxies are often in the process of falling into a cluster or have recently undergone this process, making them less affected by environmental effects compared to those residing in the denser cluster core. 
On the other hand, fully consistent with expectations, the average spectrum of the red galaxies (in red, left panel of Fig. \ref{fig:col_vs_bluredandmag}) shows several distinct absorption features, especially the sharply defined CaII$_H$ and CaII$_K$ absorption lines and Gband, that can be observed typically in old and more evolved stellar systems. This red population can serve as an efficient tracer of the cosmic web, starting from the filamentary structures that correspond to the areas where the cluster is accreting smaller systems, such as other galaxies or groups.
In summary, from the qualitative analysis of the spectral lines emerging from our stacked spectra, we can distinguish the different properties described above and reconstruct the general color bimodality behavior of the galaxies. We also highlighted, in each plot in this work, with a shaded vertical gray area, the spectral region corresponding to the joining between the blue and red arms. This region includes the H$\delta$ line, which is therefore excluded from our analysis, both in emission and in absorption. A detailed study of this feature will be presented in a forthcoming paper (Ragusa et al., in prep.), as an accurate flux calibration is required in that wavelength range.

\begin{figure}
\includegraphics[width=8.97cm]{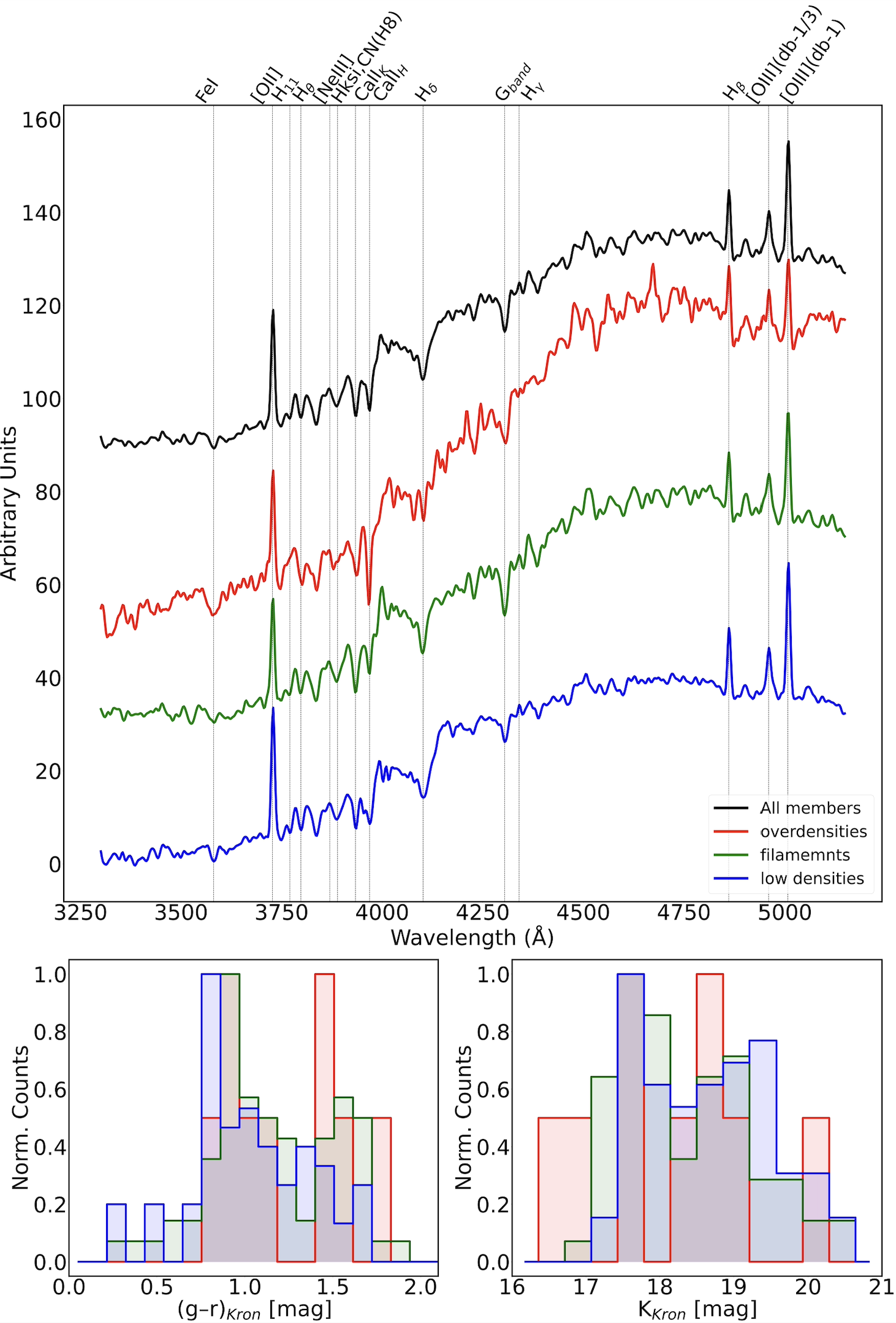}
\caption {\it Average spectral properties of our sample as a function of the local environment.} Top panel: The stacked spectrum of the AAOmega members (148) is shown in black. The stacked spectra of galaxies located in overdense regions are shown in red, those in filamentary regions in green, and those in less dense areas, comparable to the field, in blue.
All spectra are rest-frame and have been smoothed using a Gaussian kernel with a sigma of 2 pixels.  The vertical dashed gray area corresponds to the spectral region of the join between the two arms. Bottom panels: the $(g-r)_{\mathrm{Kron}}$ (left) and the $K_{\mathrm{Kron}}$ (right) distributions of the galaxies in our sample as a function of the same three (red, green and blue) $\delta$ bins.
\label{fig:delta_env}
\end{figure}
\begin{figure*}
    \includegraphics[width=9.15cm]{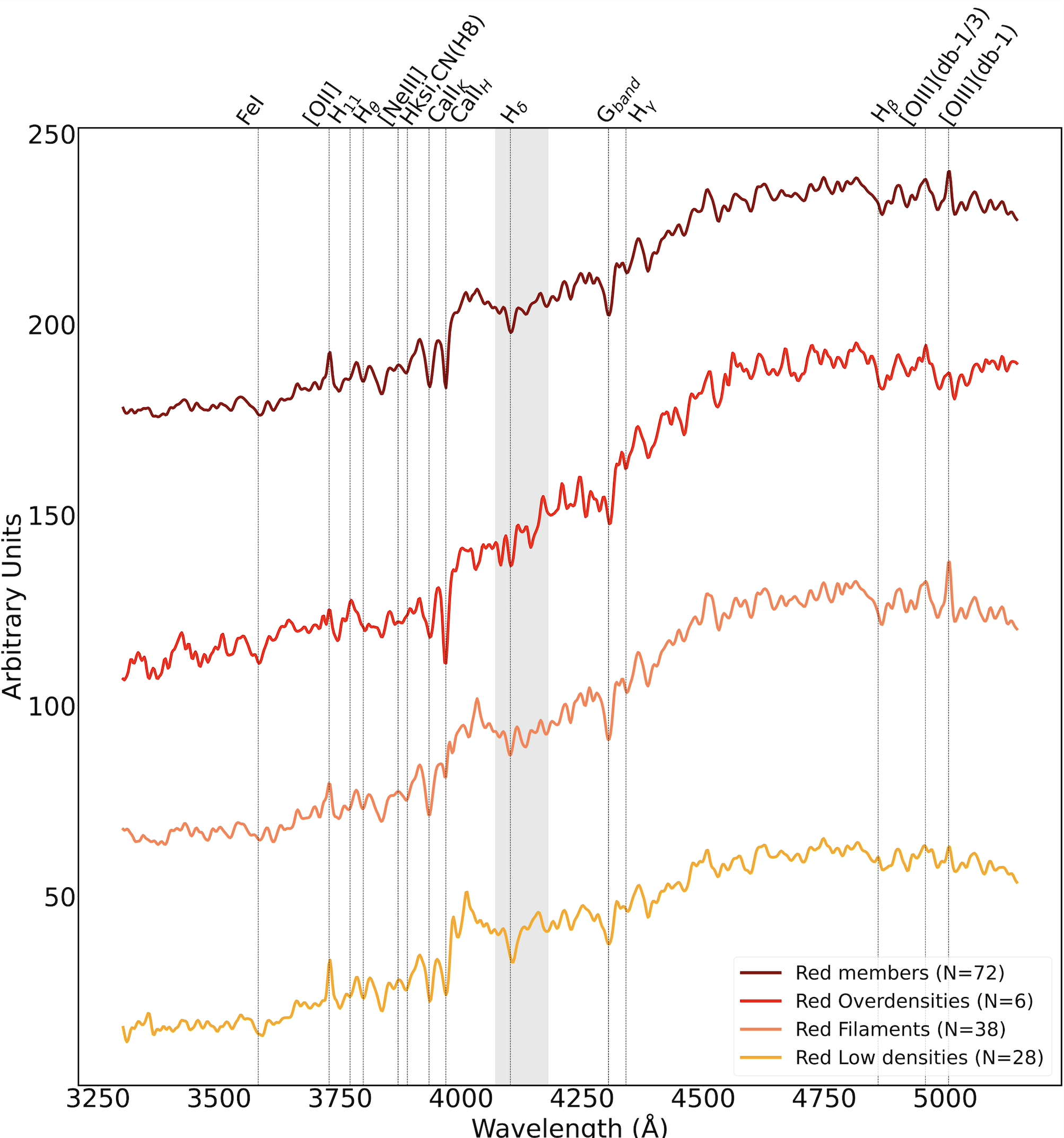}
    \includegraphics[width=9.15cm]{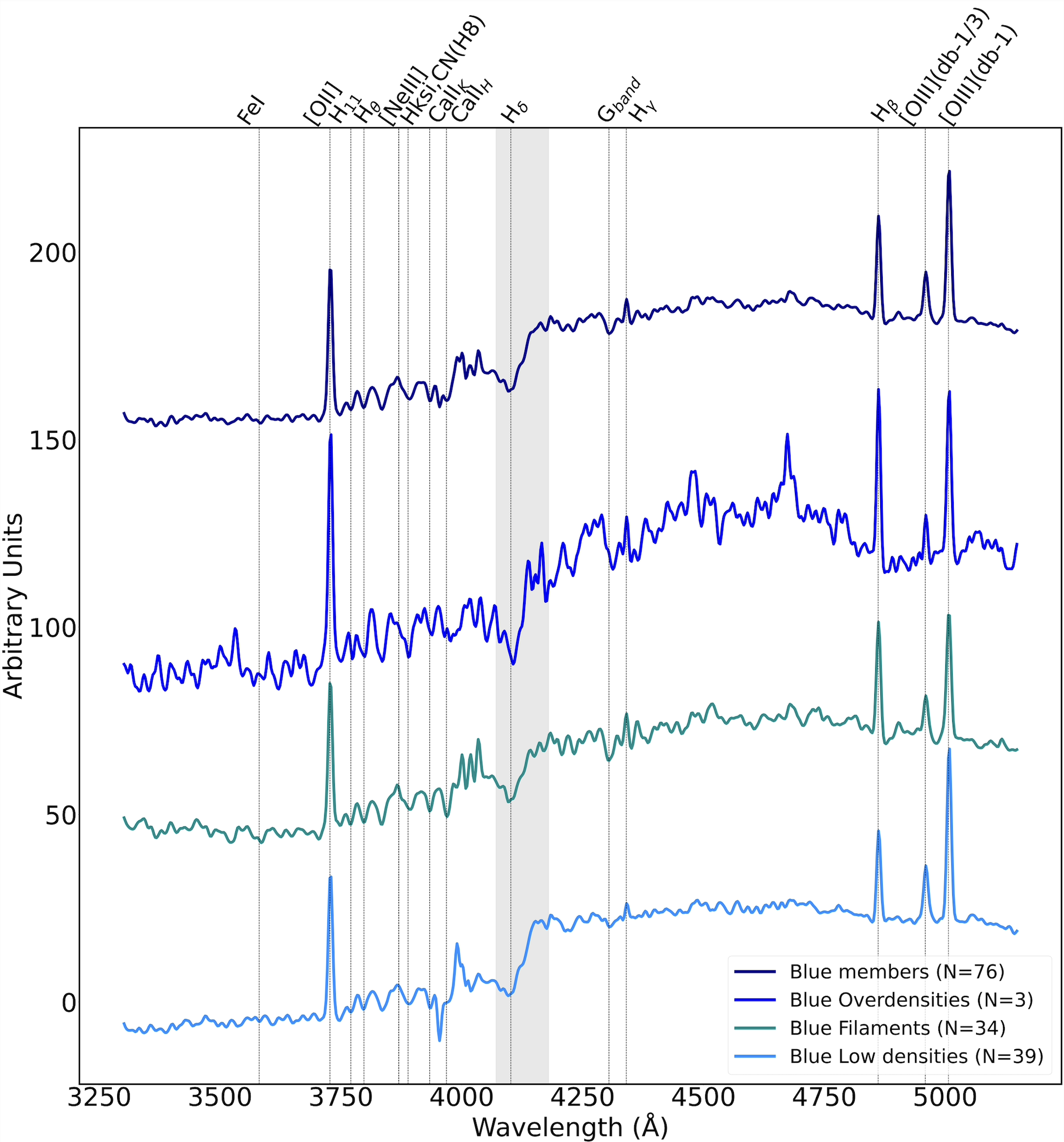}
    \caption{ {\it Average spectral properties of our sample of red and blue galaxies as a function of the local environment.} Left panel: the stacked spectrum of the population of red members (72) of MACS0416, is shown in dark red, at the top of the image. The entire sample was then divided into three subclasses based on the three bins of density $\delta$. The stacked spectra of the red galaxies located in overdense regions are shown in red, those in filamentary regions in coral, and those in less dense areas, comparable to the field, in orange. Right panel: same as the left panel, for the blue members (76) of MACS0416.
    All spectra are rest-frame and have been smoothed using a Gaussian kernel with a sigma of 2 pixels. The vertical dashed gray area corresponds to the spectral region of the join between the two arms. We performed a Kolmogorov–Smirnov (KS) test on each pair of average spectra, both in the left and right panels, obtaining p-values much smaller than 0.05 (effectively consistent with zero), which strongly supports our assumption that the average spectra of the different populations are statistically distinct.}
    \label{fig:redblue_vs_env}
\end{figure*}

\subsection{Mass dependence of spectral properties
}\label{kmag}

It is well known that galaxy mass strongly affects its evolution \citep[e.g.,][]{thomas2010MNRAS.404.1775T}. In this Section, we discuss the spectral properties of our sample galaxies as a function of their K-band luminosity used as a proxy of their stellar mass.
The K-band magnitude is a widely used proxy for estimating the stellar mass of galaxies \citep[e.g.,][]{Gavazzi1996A&A...312..397G,Kauffmann1998MNRAS.297L..23K,Sureshkumar2021A&A...653A..35S}. This is primarily because there is a relatively tight correlation between K-band luminosity and stellar mass, assuming a reasonable stellar mass-to-light (M/L) ratio \citep[e.g.][]{bell2001ASPC..230..555D}. 
Moreover the light in the near-infrared (NIR) region is less affected by dust extinction compared to optical bands, 
and more directly traces the light emitted by older, more evolved stars, particularly red giants, allowing for a more accurate assessment of stellar luminosity \citep[e.g.][]{Verheijen2001ApJ...563..694V,Kauffmann2003MNRAS.341...33K,Sureshkumar2023A&A...669A..27S}. 
Our objective here is to use the K-band magnitude within the Kron radius (K$_{\mathrm{Kron}}$), as provided by \citet{estrada2023A&A...671A.146E}, to classify the cluster member galaxies within our sample as bright or faint. 
To this aim, we divided our sample into three bins based on K$_{\mathrm{Kron}}$, defined as follows: i) K$_{\mathrm{Kron}} \leq 17.9 $ mag
(48 galaxies), ii) $17.9$ mag < K$_{\mathrm{Kron}}$ < 18.9 mag (51 galaxies) and iii) K$_{\mathrm{Kron}} \geq 18.9 $ mag (49 galaxies). These bins were specifically chosen to ensure a uniform distribution of galaxies across each category, facilitating a balanced comparison of their spectral properties across different magnitude (mass) ranges, taking into account that we selected these three K$_{\mathrm{Kron}}$ bins with similar widths within our sample to ensure a consistent level of completeness across them.
The right panel of Fig. \ref{fig:col_vs_bluredandmag} illustrates the rest-frame stacked spectra for each defined subsample,
smoothed with a Gaussian kernel with a sigma of 2 pixels. From top to bottom, it includes the mean spectrum of the entire population of cluster members (in black), followed by the spectra of the bright galaxy population (in red), the intermediate-luminosity galaxy population (in green), and finally, the faintest galaxy population (in blue). Moreover, an inset showing the trend between $(g-r)_{\mathrm{Kron}}$ and K$_{\mathrm{Kron}}$ magnitude is reported in the top left corner, confirming that, as expected, redder colors correspond to brighter and, therefore, more massive galaxies.
From a purely qualitative point of view, the spectral properties of galaxies in our sample vary systematically with K$_{\mathrm{Kron}}$ (so with the stellar mass). High-mass galaxies (red spectrum) exhibit strong absorption features (see Sec. \ref{redvsblu}, for details) with a relatively faint emission lines (such as the [OII] line), suggesting more probably an incomplete quenching scenario.
The spectrum of intermediate-mass galaxies (green) shows both emission (i.e., [OII], [OIII], H$_{\beta}$) and  absorption (i.e., CaII doublet, Gband, H$_{\delta}$) features, suggesting a transitional population with ongoing star formation.  Of particular interest is the observed inverted ratio, in the green spectrum, between the two CaII lines compared to that observed in the red one, typical of galaxies that have recently stopped their star formation activity. Indeed,  
for galaxies dominated by older stars, the $\rm{CaII}_{\rm{K}}$ line typically shows a stronger absorption than the $\rm{CaII}_{\rm{H}}$ line, resulting in a H:K ratio less than 1, as demonstrated in studies such as \citet{rose1984AJ.....89.1238R,rose1985AJ.....90.1927R}. The H:K ratio is particularly sensitive to the presence of stellar populations younger than $\sim 200~\rm{My}$ \citep{Moresco2018ApJ...868...84M}, since young stars produce absorption in H$_{\epsilon}$ which overlaps with $\rm{CaII}_{\rm{H}}$. 
Finally, the spectrum of low-mass galaxies (blue spectrum) is dominated by strong emission lines, particularly [O II], [O III],  H$_{\beta}$, H$_{\gamma}$ and H$_{\epsilon}$, the latter filling in the CaII$_H$ absorption, indicating active star formation and younger stellar populations. 
In general, as highlighted by the average spectrum of the entire sample (black spectrum), the most prominent feature is that the overall galaxy population exhibits characteristics of relatively younger stellar populations compared to the old galaxies typically found in the cores of the clusters. This suggests that, given the dataset is focused on the outskirts of the cluster, it likely includes galaxies that have not yet been fully quenched by the environment. These galaxies are presumably in a phase where they are still experiencing star formation or have only recently ceased their star-forming activity.
These results support the idea that dividing our sample based on K-band magnitude, and, by extension, galaxy stellar mass, enables the identification of distinct stellar populations. Notably, the less luminous subsample exhibits features indicative of younger stellar populations. This aligns with the expected behavior of galaxies in cluster outskirts, which are often not yet quenched and may experience ongoing or recent star formation bursts, as observed in post-starburst systems.
The spectral analysis highlighted in this study provides qualitative insights into the connection between galaxy stellar population properties and their mass, and we defer the detailed analysis of the spectral indicators in a quantitative way to the upcoming paper (Ragusa et al., in prep.). This follow-up analysis will aim to elucidate the mechanisms governing galaxy evolution within the whole cluster environment.

\subsection{Galaxies at the Edge: environmental dependence of galaxy properties}\label{Overdens}

In this Section, we will analyze the spectral properties of galaxies in the outskirts of the galaxy cluster
MACS0416, following the environmental density definition by \citet{estrada2023A&A...671A.146E}, who found three overdensity regions in the cluster periphery, and also evidenced that the properties of galaxies vary as the field density changes.
As a preliminary step in our analysis of overdensities in the outskirts, used as tracers of galaxy groups undergoing infall into the cluster, in the left panel of Fig. \ref{fig:overdensities} we superimpose the whole AAOmega dataset on the density field of objects dentified as cluster members through photometric SED fitting  \citet{estrada2023A&A...671A.146E}. The light blue dots correspond to spectroscopically confirmed members, while the white ones represent sources observed during the AAOmega spectroscopic runs but later found not to belong to the cluster. In the right panel of Fig. \ref{fig:overdensities} we have highlighted the overdensity regions A, B, and C photometrically identified by \citet{estrada2023A&A...671A.146E}, including the filamentary structures, and overlaid our spectroscopically confirmed members, color-coded according to the density of the environment in which they reside: red for overdense regions, green for filamentary regions, and blue for underdense areas.
As suggested by the right panel of Fig. \ref{fig:overdensities}, the spectroscopic follow-up appears to trace very well the overdense region B and the entire filamentary structure connecting region B to region C. Interestingly, the galaxies that trace the region B also constitute the secondary peak observed in the redshift distribution (see panel 4b in Fig.\ref{fig:histogram}) around z $\sim$ 0.385. Similarly, those tracing the filamentary structures surrounding region C appear to contribute to another secondary peak at z $\sim$ 0.405. These findings support the hypothesis that such structures may correspond to real physical systems currently experiencing infall into the main body of the cluster. The region A, on the other hand, does not seem to have a well-defined spectroscopically confirmed counterpart, although it exhibits the highest photometric density. This is primarily because the photometric member candidates in this region did not meet the magnitude limit imposed by our spectroscopic sample selection, particularly in its more crowded inner area.
In total, we spectroscopically confirm 81 galaxies in the three A, B and C overdensities plus filamentary regions. To characterize the local environment, we divided the sample of confirmed members into three distinct bins based on local density information (for details on how the mean density $\delta$ was calculated, refer to Sec. 5.2 of \citealt{estrada2023A&A...671A.146E}): low ($\delta$ $\leq$ 0), medium (0 < $\delta$ $\leq$ 1), and high ($\delta$ > 1) density. The first interval represents underdense regions, where the local density is below the mean value across the observed FoV. The second interval corresponds to overdense regions with local densities ranging up to twice the mean value (corresponding to the filamentary regions), while the third interval identifies the regions with the highest density values observed within the field (i.e., the overdensities).
In Fig. \ref{fig:delta_env}, we show the stacked spectra of all the objects that fall into the specific bin of overdensity smoothed with a Gaussian kernel with a sigma of 2 pixels: i) in red is the average spectrum for the galaxies within the overdensities ($\delta$ > 1); ii) in green that of galaxies located into the filamentary regions(0 < $\delta$ $\leq$ 1); iii) in blue that of the galaxies located in under dense regions ($\delta$ $\leq$ 0).
As usual, in black we insert the stacked spectrum of all the members. Moreover, the bottom panels show on the left the $(g-r)_{\mathrm{Kron}}$ and of the right the $K_{\mathrm{Kron}}$ distributions for each $\delta$ class (i.e. -1, 0, 1), confirming that on average, as expected, redder colors correspond to galaxies inhabiting denser local environments and that the $K_{\mathrm{Kron}}$  distributions in the three $\delta$ bins are fully consistent.
Following the same (g-r)$_{\mathrm{Kron}}$ color cut defined in Sec. \ref{redvsblu}, we divided the sample into red and blue galaxies. To analyze the global spectral behavior of red and blue galaxies across varying local environments, for each subclass, we plot their spectral properties as a function of the local density, consistent with the scheme described in Fig. \ref{fig:delta_env}. Specifically, the left panel of Fig. \ref{fig:redblue_vs_env} presents the smoothed stacked spectrum (using a Gaussian kernel with a sigma of 2 pixels) of the red population as a function of $\delta$ where a progressively darker red color corresponds to increasing $\delta$. Similarly, the right panel of Fig. \ref{fig:redblue_vs_env} displays the corresponding trend for blue galaxies, where the intensity of blue shading increases with rising $\delta$. 
Additionally, at the top of each panel, we include, for comparative purposes, the stacked and smoothed spectrum of the total population of red galaxies and blue galaxies.
As anticipated and extensively discussed in Sec. \ref{redvsblu} and Sec. \ref{kmag}, Figs. \ref{fig:delta_env} and \ref{fig:redblue_vs_env} clearly show that, as the local density increases, galaxies tend to become progressively redder, more massive, and more quenched. This trend is observed even though our sample mainly consists of galaxies located in the outskirts of the cluster  (i.e., out of the 148 member galaxies, 145 are located beyond 2R$_{200}$ ), where environmental quenching is not yet fully completed. Evidence of ongoing star formation is also apparent in the stacked spectrum of galaxies in the densest areas (see Fig. \ref{fig:delta_env}), where non-negligible emission lines persist.
Specifically, the left panel of Fig. \ref{fig:redblue_vs_env} shows that red galaxies in overdense regions exhibit more passive behavior than their lower densities counterparts, likely indicating a more evolved stellar population. This is evident from the higher H:K ratio, and the more pronounced Balmer break around 4000 $\AA$ observed in the transition from lower densities red galaxies (yellow spectrum in the left panel of Fig. \ref{fig:redblue_vs_env}) to those in the overdensities (red spectrum in the left panel of Fig. \ref{fig:redblue_vs_env}). Conversely, a younger population of blue galaxies, which are less massive and significantly more active, predominantly occupies regions of lower local density (blue spectrum in Fig. \ref{fig:delta_env}). This supports the idea that galaxy quenching is primarily driven by environmental effects, being more pronounced in areas of higher density.
Notably, when analyzing only the blue galaxy sample (right panel of Fig. \ref{fig:redblue_vs_env}), their spectral properties show minimal variation as a function of the environmental density. This homogeneity suggests that these galaxies likely represent a younger population that has undergone a relatively recent infall and has not yet had sufficient time to experience the full effects of the cluster environment. As expected, the number of blue galaxies in the high-density regions is significantly low (N=3). It is worth mentioning that we also performed Kolmogorov–Smirnov tests between each pair of K-band magnitude distributions for the galaxy populations shown in both the left and right panels of Fig. \ref{fig:redblue_vs_env}. In this case, we obtained p-values significantly greater than 0.05 (on average around 0.5), confirming that the K-band magnitude distributions of the different populations are statistically compatible, as already shown in the bottom right panel of Fig. \ref{fig:delta_env}. This supports the idea that the spectral differences shown in Fig. \ref{fig:redblue_vs_env} are not primarily driven by stellar mass, but rather reflect genuine environmental effects.
In summary, we spectroscopically confirm 81 out of 148 galaxies belonging to regions identified by \citet{estrada2023A&A...671A.146E} as filamentary or overdense areas. These galaxies are likely to have either recently experienced infall into the cluster (overdense regions) or are currently undergoing infall (filamentary regions), either as individual galaxies or as part of galaxy groups. 
As extensively discussed in Sec. \ref{kmag}, mass is also a fundamental driver for quenching mechanisms. Separating the effects of mass and environment will be the focus of a forthcoming paper (Ragusa et al., in prep.), as it will require combining the VIMOS, MUSE and AAOMEGA datasets, which will together provide a total of over 1000 members, allowing us to analyze these effects and the physical properties of the different galaxies' populations in a statistically complete way.

\section{Conclusions}\label{conclusions}

In this study, we have provided a comprehensive analysis of the outskirts of the galaxy cluster MACS0416, extending our investigation to a remarkable distance of  $ \sim$5.5$R_{200}$ (i.e., $ \sim$10 Mpc). Building on the photometric data previously obtained by \citet{estrada2023A&A...671A.146E}, we have complemented this dataset with newly acquired spectroscopic observations using the AAOmega multi-object spectrograph. The large FoV and multiplexing capabilities of AAOmega allowed us to obtain robustly estimated redshifts for 1236 unique objects inhabiting the FoV of the cluster outskirts. This unprecedented dataset has facilitated a detailed exploration of the environmental processes influencing galaxy evolution in these peripheral regions.  This work extends and complements the spectroscopic catalog previously obtained using MUSE and VIMOS data, which covered regions up to 2$R_{200}$ \citep{Ebeling2014ApJS..211...21E,Balestra2016ApJS..224...33B,Caminha2017A&A...600A..90C}.
From the spectroscopic analysis, we have constructed a catalog of redshifts for all observed objects, identifying 148 as cluster members in the outskirts of MACS0416. These spectroscopic members provide critical insights into the environmental and evolutionary processes at play. By leveraging this dataset, we analyzed the qualitative trends in galaxy spectral properties as a function of key parameters: the color, the
K-band magnitude (used as a mass proxy) and the local environmental density. This investigation revealed clear correlations between the environment and galaxy properties, particularly highlighting the significant role of overdensities and filamentary regions.
A key finding of our work is the identification of 81 galaxies associated with both overdensities and filamentary structures previously identified by \citet{estrada2023A&A...671A.146E}. These results strongly support the pre-processing scenario, where infall through the filaments of pre-assembled substructures, such as galaxy groups, plays a dominant role in the mass assembly history of the cluster. Galaxies in high-density environments exhibit spectral characteristics indicative of more massive, red, and passive populations, with colors comparable to those of the core galaxies \citep{estrada2023A&A...671A.146E}. In contrast, galaxies in low-density environments tend to be bluer, less massive, and more actively star-forming. These results highlight the transformative role of the environment, particularly in regions of higher density, where quenching mechanisms are more pronounced. Moreover, these findings underscore the importance of pre-processing phenomena in shaping galaxy properties before they enter into the cluster influence and so contribute significantly to the mass assembly and star formation histories of cluster galaxies. This is evident from the presence of evolved stellar populations in the overdense regions, where galaxies show enhanced spectral features such as the Balmer break and Ca$_H$ and Ca$_K$ absorption lines. Conversely, the spectral properties of blue galaxies remain relatively uniform across environments, suggesting a younger, less evolved population that has not yet been fully subjected to environmental effects, likely due to a more recent infall.
The findings of this study are a preliminary but crucial step in advancing our understanding of the role of the environment in galaxy evolution. The work presented here will be complemented by a forthcoming detailed and quantitative analysis of the stellar populations of these galaxies (Ragusa et al., in prep.). 

\bibliographystyle{aa}
\bibliography{MACS0416.bib}
\begin{appendix} 

\section{}\label{appA}

The raw, unreduced data acquired each night are organized into separate folders, containing the raw bias frames, raw dark frames, raw dome flat field frames, and raw science frames for both the blue and red arms. Additionally, within the raw science frames folder, for each night and each pointing, there are basic files needed to reduce the AAOmega data with a semi-automatic dedicated pipeline (see below): i) fiber flat field files indicated by the class Multi-Fibre Fibre Flat Field (MFFFF) in the AAOmega pipeline "Reduction Procedure and Configuration", which are used in the automatic reduction process; ii) MFARC files, i.e. arc exposures made with lamps having various known emission lines and used to calibrate the central wavelength and dispersion; iii) MFOBJECT, i.e. the science frames. Dome flats, which are additional files that can be used for a second order of correction, should not be confused with fiber flats and should not be used in the reduction process.
The entire data reduction process is managed by a semi-automated pipeline specifically designed for producing reduced data from {\it 2dF + AAOmega} observations. This pipeline, {\it drcontrol}, is equipped with a graphical user interface (GUI) that allows users to carefully select and handle all the reduction stages and parameters suitable for each step of the data reduction, tailored to specific scientific objectives. 
Below, we will outline the sequence of steps used in this work, aiming to produce the highest quality reduced dataset.

It is important to take into account that the data reduction process is performed separately for the two blue and red arms, which are subsequently combined at the end of the entire reduction procedure. This final combination step is also included within the pipeline, as we will discuss below.

Through an example, the steps used for the blue arm data will be presented below, noting that the same procedure was then performed on the raw data from the red arm.
In the main window of the GUI connected to the drcontrol pipeline, the recognized files in the working directory, their class, and their reduction status are displayed at each step of the data reduction described below.

\begin{itemize}
    \item {\it Create the Master Bias file}: Move into the directory containing the bias frames and run {\it drcontrol} to automatically produce the combined Master Bias file, which will then be moved to the raw science data folder and used to reduce the data. 
    
    It is crucial to keep in mind that the Master Bias file produced for the blue channel cannot be used for the red channel, which must be created separately;\\
    \item {\it Create the Master Dark file}: Copy the Master Bias created in the previous step into the directory containing the Dark frames. Since the Master Bias reflects the noise introduced by the electronics, it tends to remain stable over short periods. Therefore, if there are no current Master Bias frames available, the Master Bias file can be used (from the same channel blue or red) that was created on a previous night. Run {\it drcontrol} to automatically produce the combined Master Dark file, where the Master Bias file produced in the previous step is used in the reduction to be subtracted. 
    
    Unlike the Master Bias frame, the Master Dark file cannot be reused since it reflects the thermal noise of the detector;\\
    
    \item {\it Reducing the Science Frames}: Copy the Master Bias and the Master Dark files created in the previous steps into the directory containing the raw science data. Then, together with the MFFFF and MFARC (described above) you now have all the setup to be able to reduce automatically the raw science frames (MFOBJECTs), enabling in the GUI the Bias, Dark and Flat calibration, using a 5th-order polynomial. 
    
    Within the GUI, you can not only set up the rejection of cosmic rays and the correction for observer velocity, but also model the GAUSSIAN PSF and subtract scattered light.
    
    The sky subtraction is managed by the pipeline using the Principal Component Analysis (PCA) method on the fibers dedicated to the sky (unfortunately, the number of fibers allocated for the sky in the instrument is only 8, which does not allow for a precise and comprehensive characterization of the sky across the entire  FoV. For the sky subtraction, it is possible to configure the iterative method in the GUI, correct for telluric absorption (we set the minimum S/N to 5), choose the number of eigenvectors (in our case, a good compromise is represented by the value 100), and also define the wavelength ranges for the two channels, which in this work were 4500 - 5600 $\AA$ for the blue channel and 5800 - 8450 $\AA$ for the red channel.
   
    At the end, all the reduced science frames of the multiple observations regarding the same night and the same channel will be combined, resulting in a single .fits file containing the spectra of all the objects observed in that channel on that night;\\

    \item {\it Reducing the Dome Flats}: Dome flats are reduced in the same way as the science frames. They may be used to calibrate combined science frames, but since the performance of fibers flat is optimal, dome flats are not necessary for the aim of this work;\\

    \item{\it Splicing Red and Blue Arms Together}: Following all the previous steps, we will obtain a single combined .fits file for both the blue and the red arm for each observation set containing all the observed objects on that night. 
    
    After you have reduced both arms, you can join the two .fits together by using a dedicated tool of the pipeline, covering the entire available wavelength range.
\end{itemize}

For further details on how the {\it drcontrol} pipeline works, please refer to the official AAOmega page at the following link \url{https://aat.anu.edu.au/science/instruments/current/AAOmega/manual}, where you will find the manual with a comprehensive description of all the steps performed.

\begin{figure*}
   \includegraphics[width=18.4cm]{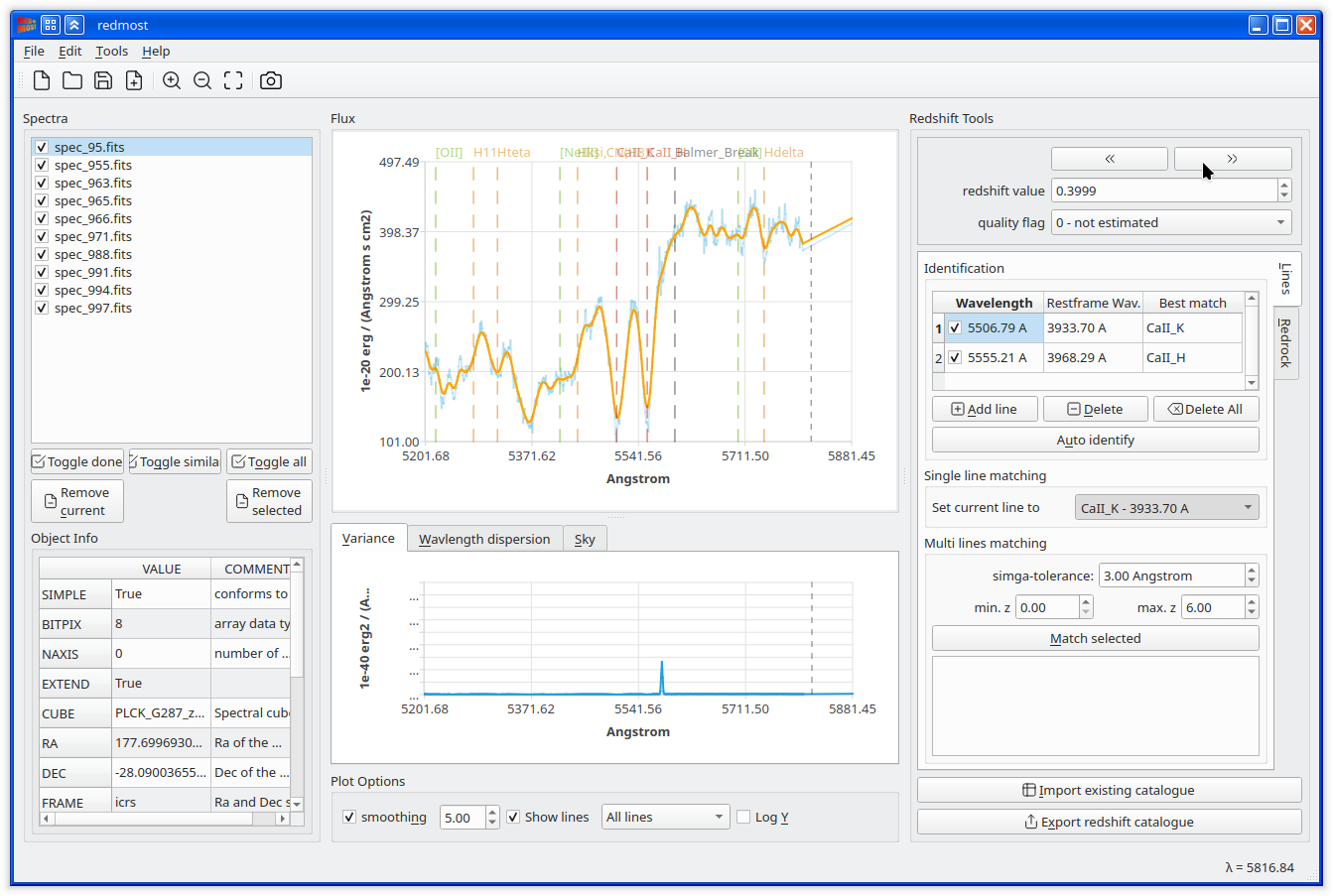}
    \caption{The main window of {\it Redmost}.}
    \label{fig:readmost_mwnd}
\end{figure*}

\begin{figure*}
   \includegraphics[width=18.4cm]{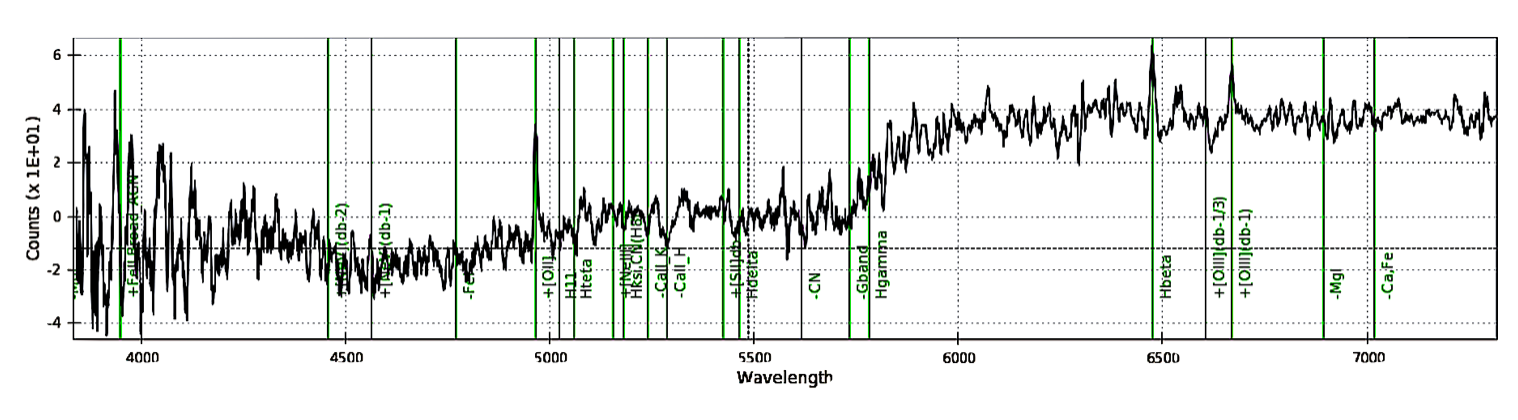}
    \caption{AAOmega spectrum of one galaxy from our sample (ID: MACS$\_$2141, z$\sim$0.33), with the redshift estimation performed using {\it EZ}. The spectrum has been smoothed with a Gaussian kernel with a sigma of 2 pixels. Key emission and absorption lines are marked with green lines.}
    \label{fig:PANDORA_SPECTRA}
\end{figure*}

\section{}\label{appB}

In this appendix, we present in detail the two independent methods used for determining the redshifts of our sample, and additionally, we introduce for the first time a new method, Redmost, developed during this work.

\begin{itemize}
    \item {\it Redrock\footnote{\url{https://github.com/desihub/redrock}} spectral template–redshift fitting code}: This tool utilizes a collection of templates to characterize the spectral features of the main object categories observed in DESI survey data \citep{levi2013arXiv1308.0847L}: quasars, galaxies, and stars. A total of 48 combinations of these spectral templates were created to offer composite solutions. Redrock finds the most accurate redshift and template matches for each spectrum by searching the entire redshift–template parameter space, choosing the best-fitting solutions based on the lowest reduced $\chi ^2$ values. 
     A notable strength of Redrock lies in its use of PCA, which allows for more flexible spectral fitting by decomposing observed spectra into a combination of principal components. This approach enhances the tool's ability to handle spectra that deviate from standard templates, thus improving its performance when fitting galaxies with complex star formation histories or atypical spectral features. 
    For further details on the analysis performed by Redrock we refer to \citet{guy2023AJ....165..144G}.
    We present here a summary of relevant literature, specifically studies on the Survey Validation of the DESI collaboration, regarding Redrock's performance and expected success rate:
    i) \citet{hahn2023AJ....165..253H} achieves a redshift efficiency of $ \sim 98.5 \%$ and a purity of $ \sim 99.5 \%$ for the faint sample of the Bright Galaxy Survey, and even better for the brighter ones; similar results were also obtained by ii) \citet{Zhou2023AJ....165...58Z} for a sample of Luminous Red Galaxies (LRGs) and by iii) \citet{Raichoor2023AJ....165..126R} in their study of the Emission Line Galaxies (ELGs). 
    Our results regarding the redshift success rate estimated by Redrock for our galaxies sample will be reported in Sec. \ref{catalogs}.
    Despite its advantages, Redrock does face some limitations. The accuracy of its redshift estimates is closely tied to the quality and completeness of the template library. Additionally, Redrock, like other automated tools, can struggle with low signal-to-noise spectra or those contaminated by skylines or other distortions, which can result in incorrect or uncertain redshift fits. 
    Another limitation lies in the treatment of stellar spectra. While Redrock includes stellar templates and is capable of fitting both extragalactic and stellar spectra, its performance on stellar objects is not as finely tuned as it is for galaxies and quasars. This can lead to occasional misclassifications, particularly in complex regions of the color-magnitude diagram or for spectra with low-quality data. 
    Moreover, the most significant limitation of Redrock is that its galaxies' templates are limited to redshift 1.6. Therefore, for objects beyond this redshift, the estimation will no longer be reliable. In such cases, complementary tools or human intervention and manual inspection may still be necessary to ensure accurate results.
    For these reasons, we adopted the complementary method described below, which enabled us to verify the accuracy of Redrock's redshift estimates through a graphical interface. In cases where the reliability was low, we were able to manually intervene in the redshift estimation process by using templates that remain reliable at redshifts greater than 1.6.

    \item {\it Easy-Z (EZ) }: {\it EZ}\footnote{\url{https://pandora.lambrate.inaf.it/EZ/}} is a software tool designed to automate the measurement of redshifts from astronomical spectra, aimed at improving the reliability of redshift determinations in large spectroscopic surveys. Developed by \citet{Garilli2010PASP..122..827G}, {\it EZ } addresses the challenge of processing vast quantities of spectral data by providing an efficient, user-friendly solution for astronomers. The software is particularly optimized for the analysis of extragalactic spectra, such as those from galaxies, quasars, and active galactic nuclei (AGN), though it can also be applied to stellar spectra in specific cases.
    The core functionality of  {\it EZ } is based on cross-correlation techniques that match observed spectra with a set of template spectra, which represent a range of known astrophysical sources. By comparing the observed spectral features (both emission and absorption lines) with these theoretical or empirical templates,  {\it EZ } determines the redshift by identifying the wavelength shift that yields the best match between the observed and template spectra. The cross-correlation process is performed using a Fourier-based algorithm, which significantly enhances computational efficiency, making it suitable for the analysis of large datasets such as those produced by contemporary redshift surveys.
    Moreover,  {\it EZ } is designed to be flexible and allows for manual intervention when necessary. Users can interact with the software through a GUI that enables the visualization of individual spectra, inspection of the redshift fit, and fine-tuning of the fitting parameters. This balance between automation and user control makes  {\it EZ } a powerful tool for both experienced astronomers and non-expert users.
    \end{itemize}
    \begin{itemize}
    \item   {\it Redmost}: REDshift Measurements Of SpecTra (Redmost, \citealt[][]{redmost}) is a GUI application we developed to perform redshift measurements on 1D spectra. The program is written in Python, based on the packages {\it astropy} and {\it specutils} \citep{Astropy2022}. The GUI is built with the Qt6 graphics library and it supports both PyQt6 and PySide6 backends, with precedence to PyQt6 if both are installed. PyQt5 backend is also supported for compatibility. This makes it completely cross-platform and fully compatible with most of the widely used operating systems (like GNU-Linux, Microsoft Windows, and Apple macOS). It can be installed directly using \texttt{pip}, either using a pre-packaged release or using the source code which is freely available on GitHub under the BSD-3-Clause license\footnote{\url{https://github.com/mauritiusdadd/redmost}}.\\
The main window of the program (shown in Fig. \ref{fig:readmost_mwnd}), is designed to let the users have under control all the most important information: in the central section there is an interactive plot of the spectrum they are currently working on, while on the left side some basic information is shown along with a list of the other open spectra. In the right panel, instead, there is a set of tools that can be used to measure the redshift. Currently, {\it Redmost} provides out-of-the-box two methods to measure the redshifts:

        \begin{itemize}
    \item{a line matching tool, that allows to measure the redshift by manually selecting the lines on the spectra and matching their positions with a list of known ones;}\\
    \item{a Redrock plugin that uses the functions and spectral templates provided by the \texttt{redrock} python module, which is developed and maintained by the Dark Energy Spectroscopic Instrument (DESI, \citealt{DESIREF2014}) collaboration.}
\end{itemize}

The Redrock plugin, in particular, allows to run Redrock either on all the open spectra or just on a user-selected subset and it automatically takes care to format the input data and retrieve the results. Since {\it Redmost} is built to be modular, following a plug-in style philosophy, redshift measurement functionalities can be easily extended and we will develop new plug-ins for other methods in future releases. The strength of this approach is that the redshifts can be measured independently by different methods and the results can be checked and validated directly on the spectrum plots, interactively, and within the same program.

The spectroscopic redshift measurements can be exported either as a FITS table (or any other format supported by astropy) or as a {\it Redmost} project. The latter is a JSON file that contains in plain text all the information needed to restore the status of the application, beyond the measured redshifts. {\it Redmost} is in active development and the full documentation of the latest version along with some tutorials are available on the project home page: \url{https://redmost.readthedocs.io}.

    \end{itemize}    
\end{appendix}

\begin{acknowledgements}
{\bf We sincerely thank the anonymous referee for improving the overall quality of the paper. We greatly appreciate their time and effort in reviewing our work.}
RR, MDA, AM, ML and AI acknowledge financial support from INAF Large Grant 2022, FFO 1.05.01.86.16.
ML and AI acknowledge financial contribution from the Italian Ministry grant Premiale MITIC 2017 and from INAF-Minigrant "4MOST- StePS: a Stellar Population Survey using 4MOST@VISTA" (2022).
RR, MDA and AM acknowledge financial support through grants PRIN-MIUR 2020SKSTHZ, PRIN-MIUR
2017WSCC32 (PI: P. Rosati), INAF “main-stream” 1.05.01.86.20 (PI: M. Nonino)” and INAF “main-stream” 1.05.01.86.31 (P.I.: E. Vanzella). RR, MDA and AM acknowledge \citet{Balestra2016ApJS..224...33B}, \citet{Caminha2017A&A...600A..90C} and the VLT programme IDs 186.A-0798,  094.A-0115(B), 094.A-0525(A).
AG acknowledges the financial support of the MUR through the program “Dipartimenti di Eccellenza 2018-2022” (Grant SUPER-C).

\end{acknowledgements}
\section*{Institutions}

\noindent
$^{1}$ INAF – Osservatorio Astronomico di Capodimonte, Salita Moiariello 16, 80131 Napoli, Italy.\\
$^{2}$ Università di Salerno, Dipartimento di Fisica “E.R. Caianiello”, Via Giovanni Paolo II 132, 84084 Fisciano (SA), Italy.\\
$^{3}$ Dipartimento di Fisica e Scienze della Terra, Università degli Studi di Ferrara, Via Saragat 1, 44122 Ferrara, Italy.\\
$^{4}$ INFN – Gruppo Collegato di Salerno – Sezione di Napoli, Dipartimento di Fisica “E.R. Caianiello”, Università di Salerno, Via Giovanni Paolo II, 132, 84084 Fisciano (SA), Italy.\\
$^{5}$ INAF – Osservatorio Astronomico di Brera, via Brera 28, I-20121 Milano, Italy.\\
$^{6}$ Dipartimento di Fisica dell’Università degli Studi di Trieste – Sezione di Astronomia, via Tiepolo 11, I-34143 Trieste, Italy.\\
$^{7}$ INAF – Osservatorio Astronomico di Trieste, via Tiepolo 11, I-34143 Trieste, Italy.\\
$^{8}$ Centro de Astrobiología (CAB), CSIC-INTA, Ctra. de Ajalvir km 4, Torrejón de Ardoz 28850, Madrid, Spain.\\
$^{9}$ Thales Alenia Space, Via E. Mattei 1, 20064, Gorgonzola, Italy.\\
$^{10}$ Dipartimento di Fisica, Università degli Studi di Milano, via Celoria 16, I-20133 Milano, Italy.\\
$^{11}$ INAF – IASF Milano, via A. Corti 12, I-20133 Milano, Italy.\\
$^{12}$ INAF – Osservatorio Astronomico di Padova, Vicolo dell’Osservatorio 5, I-35122 Padova, Italy.\\
$^{13}$ Dipartimento di Fisica e Astronomia "G. Galilei", Università di Padova, Via Marzolo 8, 35131 Padova, Italy.\\
$^{14}$ INAF – OAS, Osservatorio di Astrofisica e Scienza dello Spazio di Bologna, via Gobetti 93/3, I-40129 Bologna, Italy.\\
$^{15}$ Max-Planck-Institut für Physik, Boltzmannstr. 8, 85748 Garching, Germany.\\
$^{16}$ Universitäts-Sternwarte München, Fakultät für Physik, Ludwig-Maximilians-Universität München, Scheinerstr. 1, 81679 München, Germany.\\
$^{17}$ Max-Planck-Institut für Extraterrestrische Physik, Giessenbachstr. 1, 85748 Garching, Germany.\\
$^{18}$ Department of Physics "E. Pancini", University Federico II, Via Cintia 21, 80126 Napoli, Italy.\\
$^{19}$ TUM School of Natural Sciences, Technical University of Munich, Garching 85748, Germany.\\
$^{20}$ Dipartimento di Fisica e Geologia, Università degli Studi di Perugia, Via Alessandro Pascoli, s.n.c., 06123 Perugia, Italy.\\
$^{21}$ INFN Sezione di Perugia, Via Alessandro Pascoli, s.n.c., 06123 Perugia, Italy.\\
\end{document}